\documentclass[fleqn,usenatbib]{mnras}

\usepackage[T1]{fontenc}

\DeclareRobustCommand{\VAN}[3]{#2}
\let\VANthebibliography\thebibliography
\def\thebibliography{\DeclareRobustCommand{\VAN}[3]{##3}\VANthebibliography}

\usepackage{graphicx}	
\usepackage{amsmath}	
\usepackage{amssymb}	
\usepackage{booktabs}

\newcommand{\Msol}{\ensuremath{M_{\odot}}}
\newcommand{\mysun}{\ensuremath{{\odot}}}
\newcommand{\suchthat}{\;\ifnum\currentgrouptype=16 \middle\fi|\;}

\usepackage[greek, english]{babel}

\usepackage{ulem}
\newcommand{\orcid}[1]{\href{https://orcid.org/#1}{\includegraphics[width=10pt]{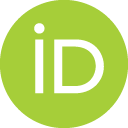}}}

\usepackage{newtxtext,newtxmath}

\title[Strong lensing mass models of MACS\,J0242.5$-$2132 \& MACS\,J0949.8$+$1708]{Joint \textit{HST}, VLT/MUSE and \textit{XMM-Newton} observations to constrain the mass distribution of the two strong lensing galaxy clusters: MACS\,J0242.5$-$2132 \& MACS\,J0949.8$+$1708}

\author[Joseph F. V. \textsc{Allingham} et al.]{
Joseph F. V. \textsc{Allingham}\orcid{0000-0003-2718-8640}$^{1}$\thanks{E-mail: \href{mailto:joseph.allingham@sydney.edu.au}{joseph.allingham@sydney.edu.au}},
Mathilde \textsc{Jauzac}\orcid{0000-0003-1974-8732}$^{2,3,4,5}$,
David J. \textsc{Lagattuta}\orcid{0000-0002-7633-2883}$^{2,3}$,
\newauthor
Guillaume \textsc{Mahler}\orcid{0000-0003-3266-2001}$^{2,3}$,
C\'eline \textsc{B\oe{}hm}\orcid{0000-0002-5074-9998}$^{1}$,
Geraint F. \textsc{Lewis}\orcid{0000-0003-3081-9319}$^{1}$,
Dominique \textsc{Eckert}\orcid{0000-0001-7917-3892}$^{6}$,
\newauthor
Alastair \textsc{Edge}\orcid{0000-0002-3398-6916}$^{2}$,
and Stefano \textsc{Ettori}\orcid{0000-0003-4117-8617}$^{7,8}$
\\
$^{1}$School of Physics, A28, The University of Sydney, NSW 2006, Australia;\\ 
$^{2}$Centre for Extragalactic Astronomy, Department of Physics, Durham University, South Road, Durham DH1 3LE, UK;\\
$^{3}$Institute for Computational Cosmology, Department of Physics, Durham University, South Road, Durham DH1 3LE, UK;\\
$^{4}$Astrophysics Research Centre, University of KwaZulu-Natal, Westville Campus, Durban 4041, South Africa;\\
$^{5}$School of Mathematics, Statistics \& Computer Science, University of KwaZulu-Natal, Westville Campus, Durban 4041, South Africa;\\
$^{6}$Department of Astronomy, University of Geneva, ch. d’Écogia 16, CH-1290 Versoix Switzerland;\\
$^{7}$INAF - Osservatorio di Astrofisica e Scienza dello Spazio di Bologna, via Piero Gobetti 93/3, 40129 Bologna;\\
$^{8}$INFN, Sezione di Bologna, viale Berti Pichat 6/2, 40127 Bologna, Italia.
}

\date{Accepted 2023 March 3. Received 2023 January 31; in original form 2022 July 20}

\pubyear{2023}

\begin{document}
\label{firstpage}
\pagerange{\pageref{firstpage}--\pageref{lastpage}}
\maketitle 

\begin{abstract}
We present the strong lensing analysis of two galaxy clusters: MACS\,J0242.5-2132 (MACS\,J0242, $z=0.313$) and MACS\,J0949.8+1708 (MACS\,J0949, $z=0.383$). 
Their total matter distributions are constrained thanks to the powerful combination of observations with the \textit{Hubble Space Telescope} and the MUSE instrument.
Using these observations, we precisely measure the redshift of six multiple image systems in MACS\,J0242, and two in MACS\,J0949. We also include four multiple image systems in the latter cluster identified in \textit{HST} imaging without MUSE redshift measurements.
For each cluster, our best-fit mass model consists of a single cluster-scale halo, and 57 (170) galaxy-scale halos for MACS\,J0242 (MACS\,J0949). 
Multiple images positions are predicted with a $rms$ 0.39\arcsec and 0.15\arcsec for MACS\,J0242 and MACS\,J0949 models respectively.
From these mass models, we derive aperture masses of $M(R<$200\,kpc$) = 1.67_{-0.05}^{+0.03}\times 10^{14}\Msol$, and $M(R<$200\,kpc$) = 2.00_{-0.20}^{+0.05}\times 10^{14}\Msol$.
Combining our analysis with X-ray observations from the \textit{XMM-Newton Observatory}, we show that MACS\,J0242 appears to be a relatively relaxed cluster, while conversely, MACS\,J0949 shows a relaxing post-merger state.
At 200\,kpc, X-ray observations suggest the hot gas fraction to be respectively $f_g = 0.115^{+0.003}_{-0.004}$ and $0.053^{+0.007}_{-0.006}$ for MACS\,J0242 and MACS\,J0949.
MACS\,J0242 being relaxed, its density profile is very well fitted by a NFW distribution, in agreement with X-ray observations.
Finally, the strong lensing analysis of MACS\,J0949 suggests a flat dark matter density distribution in the core, between 10 and 100\,kpc.
This appears consistent with X-ray observations.
\end{abstract}




\section{Introduction}

One of the most promising avenues towards understanding the nature of dark matter is to study its gravitational influence on the Universe's large-scale structure, particularly within the most massive galaxy clusters. 
These gravitationally bound clusters act as the largest natural laboratories, allowing not only to observe the large-scale baryonic physics, but also to indirectly probe dark matter thanks to the effect of gravitational lensing. Gravitational lensing is the phenomenon of optical distortion of background images, occurring when a massive foreground object -- like a cluster, the ``lens'' -- is on its line-of-sight.  
Gravitational lenses act as magnifying telescopes of objects in the background, creating in some cases multiple images of a same source, and allowing observers to study objects in the distant Universe \citep[for a review, see][]{Kneib_2011}.

For all these reasons, since the first discovery of the gravitational giant arc of Abell\,370 \citep{1987hrpg.work..467H, 1988A&A...191L..19S} to the modern surveys of galaxy clusters and gravitational lenses such as the Cluster Lensing And Supernovae survey with Hubble \citep[CLASH,][]{CLASH_reference}, the Hubble Frontier Fields \citep[HFF, PI: Lotz,][]{Lotz_2017}, the REionization LensIng Cluster Survey \citep[RELICS, PI: Coe,][]{coe_2019}, the SDSS Giant Arcs Survey  \citep[SGAS, PI: Gladders,][]{SGAS_ref} and the Beyond the Ultra-deep Frontier Fields And Legacy Observation programme \citep[BUFFALO, PI: Steinhardt \& Jauzac,][]{Steinhardt_2020}, gravitational lensing has emerged as a field of cosmology, capable of bringing key information to comprehend the structure formation and the nature of dark matter. 

In particular, the study of a system of multiple images originating from one source through gravitational lensing allows one to constrain the mass distribution within the lens, and to characterise the dark matter density profile within it. The descriptive potential of gravitational lensing has already been showcased at multiple occasions such as in \citep[][]{2014MNRAS.444..268R, Jauzac_2014b, Jauzac2016, 2015MNRAS.447.3130D, 2015MNRAS.451.3920D, 2016MNRAS.459.3447D,2018MNRAS.473.4279D, 2020ApJ...904..106D,  2015ApJ...800...38G, 2017A&A...607A..93C, 2018MNRAS.480.3140W}.
Using the combination of high resolution images taken with the \textit{Hubble Space Telescope} (\textit{HST}) and the Dark Energy Survey (DES) for photometric analysis in the one hand, and the Multi-Unit Spectroscopic Explorer \citep[MUSE, see][]{2014Msngr.157...13B} for spectroscopy in the other, we were able to securely identify cluster members and multiple images systems. This combination has proven to be particularly successful over the past few years \citep[e.g.][]{2016ApJ...817...60T, 10.1093/mnras/stx1079, 10.1093/mnras/stz620, 2016MNRAS.457.2029J, 2019MNRAS.483.3082J, 2021MNRAS.508.1206J, Grillo_2016,  Malher_2017, 2019A&A...632A..36C}.

In this paper, we repeat a similar exercise, looking at two galaxy clusters, MACS\,J0242.5$-$2132 and MACS\,J0949.8$+$1708 (i.e. RXC J0949.8+1707), hereafter MACS\,J0242 and MACS\,J0949 respectively, initially discovered by the MAssive Cluster Survey \citep[MACS, PI: Ebeling,][]{article_MACS_Ebeling_2001}. We combined multi-band \textit{HST} and ground-based imaging with spectroscopy from VLT/MUSE with the lensing modelling technique presented in detail in \citet{Richard_2014} which makes use of the publicly available \textsc{Lenstool} software \citep{1996ApJ...471..643K, 2007NJPh....9..447J}.
We then confront our lensing results to the intra-cluster gas distribution observed by the \textit{XMM-Newton X-ray Observatory}. 

It is common practice to use the combined baryonic analysis of the X-ray signal and the Sunyaev-Zel'dovich effect (SZ) to understand the thermodynamics of galaxy clusters. One can then reconstruct the total matter density of galaxy clusters by making a number of hypotheses such as hydrostatic equilibrium or polytropic temperature distribution \citep[see][]{Tchernin_2018}.
Furthermore, as the analysis of multi-wavelengths observations (optical, Sunyaev-Zel'dovich effect, X-rays) characterises the thermodynamics of the intra-cluster medium \citep[ICM; see][]{Sereno_2017}, a careful comparison between these and a strong lensing analysis can provide clues on the possible differences between expected and observed baryon and dark matter distributions.

As an example, the study in merging galaxy clusters of the offset between the position of the centre of dark matter, luminous galaxies and X-ray emission
can be used to constrain the cross-section of self-interacting dark matter \citep[SIDM, see][for an overview]{Tulin_2018}.
In fact, simulations of colliding clusters suggests the cold dark matter (CDM) distribution to be bounded to the luminous distribution; while in SIDM scenarios dark matter lags behind baryonic matter \citep{10.1111/j.1365-2966.2011.18246.x, 10.1093/mnras/stw2670, 10.1093/mnras/stx463}.
For instance, \citet{10.1093/mnras/stx463} present SIDM simulations with anisotropic scattering, yielding an offset between the galaxies centre and that of dark matter (DM) smaller than 10\,kpc for an interaction $\sigma / m = 1$\,cm$^2$.g$^{-1}$.
This was pioneered in \cite{2004ApJ...604..596C} and \cite{2008ApJ...687..959B}, and has now become more and more popular as shown in, e.g. \cite{2011MNRAS.417..333M, Harvey_2015, 2015MNRAS.449.3393M, 2018MNRAS.477..669M, 2016MNRAS.463.3876J, 2018MNRAS.477.4046J}.

In this article, we focus on the lensing-based mass reconstructions of the two clusters.
Utilising the ICM detected in the X-rays to infer the dark matter halo profile, we compare the results of our lensing reconstruction to the \textit{XMM-Newton} X-ray data from \citet{2021A&A...650A.104C}, processed following the X-COP pipeline \citep[][]{ghirardini19} for these two clusters.
We present a broader context for such comparisons, i.e. new models of baryonic matter distribution rooted in lensing analysis to constrain the electronic densities of galaxy clusters, in a companion paper (Allingham et al.\ in prep.).

Our paper is organised as follows. 
In Section \ref{sec:Data}, we present the observations used for our analysis. The methods to extract multiple image candidates, and to build cluster galaxy catalogues are presented in Section \ref{sec:Analysis}. The lensing reconstruction method is introduced in Section \ref{sec:Mass-modelling}, the mass models are described in Section \ref{sec:Results}, and conclusions are presented in Section \ref{sec:Discussion}. 
Throughout this paper,  we assume the $\Lambda$CDM cosmological model, with $\Omega_m=0.3$, $\Omega_{\Lambda}=0.7$, and $H_{0}=70$\,km/s/Mpc.
All magnitudes use the AB convention system \citep{1974ApJS...27...21O}.

\section{Data}
\label{sec:Data}

\begin{figure*}
    \centering
    \includegraphics[width=\textwidth]{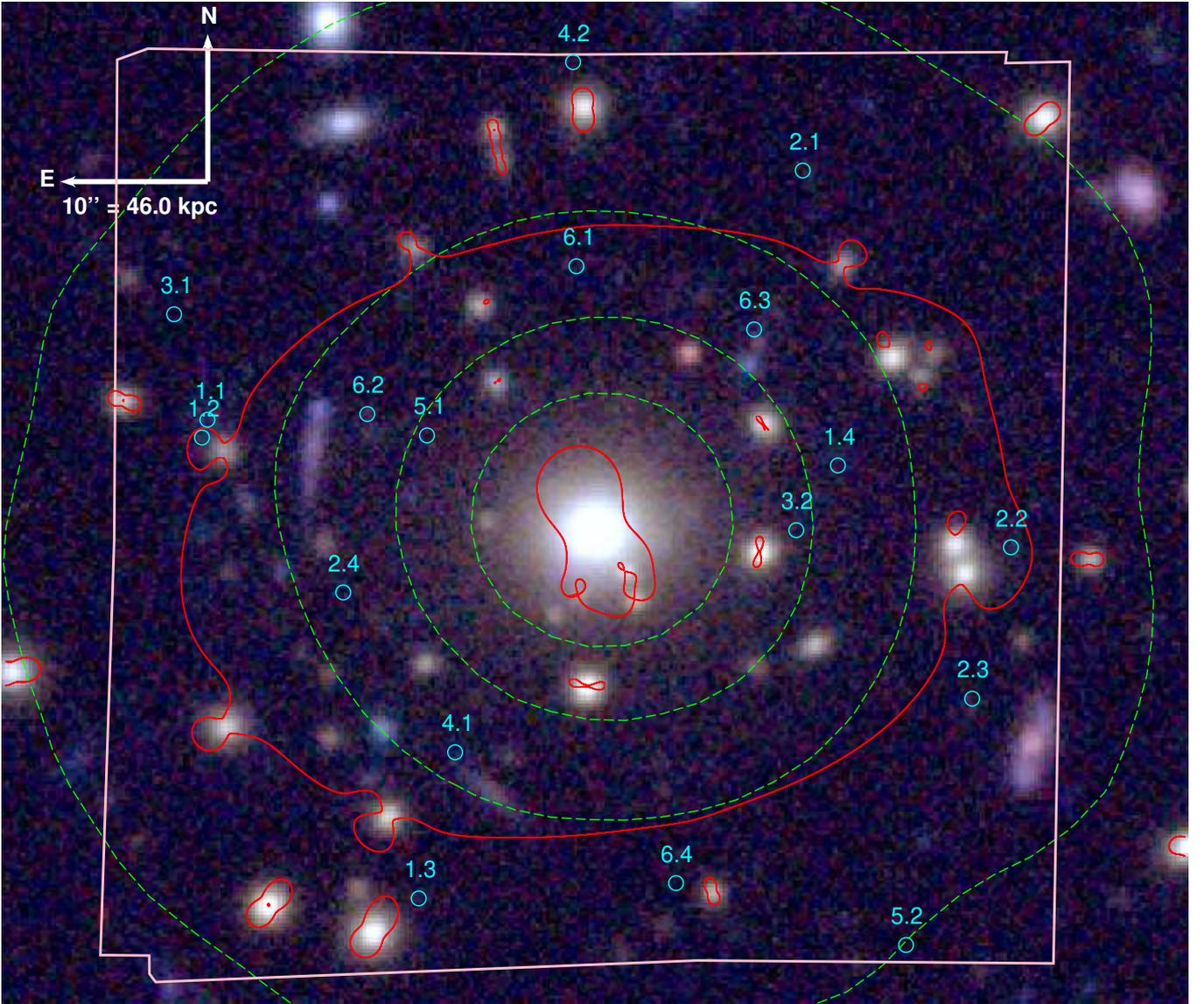}
    \caption{Composite DES colour image of MACS\,J0242. 
    The gas distribution obtained from \textit{XMM-Newton} observations is shown with dashed green contours. In cyan, we highlight the positions of the multiple images used to constrain the mass model, and which are listed in Table\,\ref{tab:spectro_mul_m0242}. Critical lines for a source at $z=3.0627$ (redshift of system 1) are shown in red.
    The MUSE field of view is shown in pink.}
    \label{fig:astro_im_m0242}
\end{figure*}

To determine the cluster mass distributions as robustly as possible, we include both imaging and spectroscopic information when constructing lens models. This combination is especially powerful, allowing us to identify and confirm individual components of the model (such as multiple-image constraints and cluster members), while simultaneously rejecting interlopers along the line of sight. 
We complement the observations we have with \textit{HST} and VLT/MUSE with \textit{XMM-Newton X-ray Observatory} observations to cross-check our lensing model results.
Figures\,\ref{fig:astro_im_m0242} and \ref{fig:astro_im_m0949} present a stack of the imaging, spectroscopic, and X-ray data for clusters MACS\,J0242 and MACS\,J0949 respectively.

\subsection{Imaging}
\subsubsection{Hubble Space Telescope}

As part of the MACS survey \citep[][]{article_MACS_Ebeling_2001}, both targets in our study have publicly available \textit{HST} data. Snapshot (1200s) imaging of MACS\,J0242 taken with the Wide Field Planetary Camera 2 \citep[WFPC2,][]{WFPC2_reference} exist for both the F606W and F814W bands (PID:11103, PI: Ebeling), supplemented by an additional 1200s image taken with the Advanced Camera for Surveys \citep[ACS,][]{1998SPIE.3356..234F} in F606W (PID: 12166, PI: Ebeling). Similarly, shallow imaging for MACS\,J0949 have been taken with the ACS in both F606W (PID:10491, PI: Ebeling) and F814W (PID: 12166, PI: Ebeling).  Archival processed versions of these datasets are available from the \textit{Hubble} Legacy Archive\footnote{\url{https://hla.stsci.edu/}}.

Following the initial MACS data, MACS\,J0949 was subsequently observed as part of the RELICS survey \citep[][]{coe_2019} -- under the name RXC J0949.8+1707 -- and thus there are additional data sets for this cluster. Specifically, ACS imaging in F435W, F606W and F814W provide wider, deeper coverage of the cluster field in optical bands, while coverage in F105W, F125W, F140W and F160W bands using the Wide Field Camera 3 \citep[WFC3,][]{WFC3_IR_photometry_reference} provide information in the near-IR regime. These data are also publicly available\footnote{\url{https://archive.stsci.edu/prepds/relics/}}, and therefore in this work combine all of the imaging (save for the F435W band, which is too low S/N for our purposes) to create our master data set. A summary of all available \textit{HST} imaging can be found in Table \ref{tab:data_HST}.

\subsubsection{DESI Legacy Survey}
Since the available \textit{HST} imaging for MACS\,J0242 are shallow and colour information is limited to a WFPC2-sized footprint, we complement these data with additional multi-band ground-based imaging from the Dark Energy Spectroscopic Instrument (DESI) \href{https://www.legacysurvey.org/}{legacy archive}. To enhance the \textit{HST} data as much as possible, we extract cutout images in three optical bands -- g, r and z, see \citet{Abbott_2018}. The images are centred around the MACS\,J0242 brightest cluster galaxy (BCG) located at ($\alpha = 40.6497\deg$, $\delta = -21.5406\deg$), and extend over a full ACS field of view. Combining the space- and ground-based information allow us to improve our galaxy selection function during lens modelling (see Section \ref{sec:Analysis}). The DESI data are summarised in Table\,\ref{tab:data_DESI}.

\begin{table*}
	\centering
	\caption{Summary of the \textit{HST} observations used in this analysis for MACS\,J0242 and MACS\,J0949.}
	\label{tab:data_HST}
	\begin{tabular}{ccccccc}
	    \hline
		\hline
		Galaxy cluster & Date of observation & Proposal & Camera/Filter & RA (°, J2000) & Dec (°, J2000) & Exposure time (s)\\
		\hline
        MACS J0242 & 29/02/2012 & 12166 & ACS/F606W & 40.645985  & -21.541129 & 1200\\
        & 30/11/2007 & 11103 & WFPC2/F606W & 40.649625 & -21.540556 & 1200\\
        & 27/10/2008 & 11103 & WFPC2/F814W & 40.649625 & -21.540556 & 1200\\
        \hline
        MACS J0949 & 09/10/2015 & 14096 & WFC3/F105W & 147.462029 & 17.120908 & 706\\
        & 09/10/2015 & 14096 & WFC3/F125W & 147.462029 & 17.120908 & 356\\
        & 09/10/2015 & 14096 & WFC3/F140W & 147.462029 & 17.120908 & 331\\
        & 09/10/2015 & 14096 & WFC3/F160W & 147.462029 & 17.120908 & 906\\
        & 20/11/2015 & 14096 & ACS/F606W & 147.463077 & 17.120878 & 1013\\
        & 20/11/2015 & 14096 & ACS/F814W & 147.463077 & 17.120878 & 1013\\
        & 23/04/2011 & 14096 & ACS/F814W & 147.463077 & 17.120878 & 1440\\
        & 25/10/2005 & 14096 & ACS/F606W & 147.463077 & 17.120878 & 1200\\
		\hline
		\hline
	\end{tabular}
\end{table*}

\begin{table*}
	\centering
	\caption{Summary of the DESI observations used in this analysis for MACS\,J0242.}
	\label{tab:data_DESI}
	\begin{tabular}{ccccccc}
	    \hline
		\hline
		Date of Observation$^a$ & Proposal & Filter & RA (°, J2000) & Dec (°, J2000) & Exposure time (s) & Seeing (\arcsec)$^a$\\
		\hline
        24/09/2016 & 2012B-0001 & DES/g & 40.6497 & -21.5406 & 810 & 0.738\\
        05/11/2016 & 2012B-0001 & DES/r & 40.6497 & -21.5406 & 720 & 0.701\\
        16/11/2016 & 2012B-0001 & DES/z & 40.6497 & -21.5406 & 810 & 0.859\\
		\hline
		\hline
	\end{tabular}
	\\
	\begin{flushleft}
	~~~~~~~~~~~~~~~~~~~~~~~~~~~~~~$^a$ Median values, determined over all observations	
	\end{flushleft}

\end{table*}

\begin{table*}
	\centering
	\caption{Summary of MUSE observations for MACS\,J0242 and MACS\,J0949. Columns 1 to 3 indicate respectively the name of the cluster, its average redshift, and the ID of the ESO programme. For each pointing, we then give the observation date in column 4, the right ascension, R.A., and declination, Dec., of the centre of the field of view in columns 5 and 6, the total exposure time in column 7, and the full width at half maximum (FWHM) of the seeing during the observations in column 8.}
	\label{tab:data_MUSE}
	\begin{tabular}{cccccccc}
	    \hline
		\hline
		Galaxy cluster & $z$ & Date of observation & ESO proposal & RA (°, J2000) & Dec (°, J2000) & Exposure time (s) & Seeing (\arcsec)\\
		\hline
        MACS J0242 & $0.3131$ & 26/12/2017 & 0100.A-0792(A) & 40.650167 & -21.5401389 & 2910 & 0.63\\
        \hline
        MACS J0949 & $0.383$ & 20/02/2020 & 0104.A-0801(A) & 147.465792 & 17.119528 & 2910 & 0.71\\
		\hline
		\hline
	\end{tabular}
\end{table*}

\begin{figure*}
    \centering
    \includegraphics[width=\textwidth]{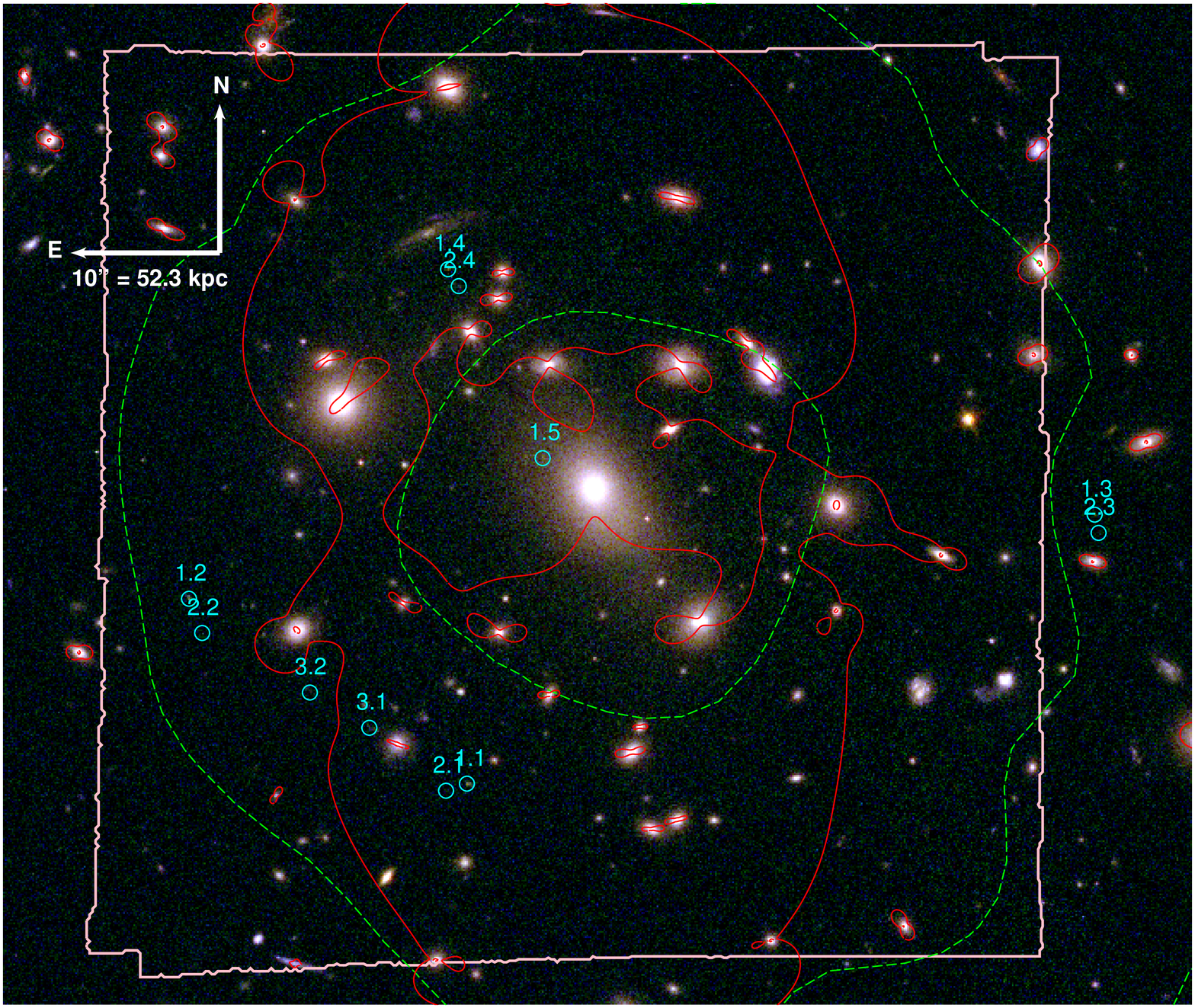}
    \caption{Composite colour \textit{HST} image of MACS\,J0949.
    The critical lines of system 1, at redshift $4.8902$, are shown in red.
    The gas distribution obtained thanks to \textit{XMM-Newton} observations are shown with dash green contours. In cyan, we highlight the positions of the multiple images used to constrain the mass model. They are listed in Table\,\ref{tab:spectro_mul_m0949}. In pink, we display the MUSE field of view.}
    \label{fig:astro_im_m0949}
\end{figure*}

\subsection{Spectroscopy}
\label{sec:Data_spectro}

In addition to imaging, our lensing reconstruction makes use of the Multi Unit Spectroscopic Explorer \citep[MUSE;][]{2014Msngr.157...13B} observations at the Very Large Telescope. Such observations are invaluable to obtain redshift information. Both clusters were observed with MUSE as part of the filler large programme ``A MUSE Survey of the Most Massive Clusters of Galaxies - the Universe's Kaleidoscopes'' (PI: Edge). 
Data for each cluster consists of a single MUSE pointing, divided in a series of three exposures of 970\,seconds. To reduce the effects of bad pixels, cosmic rays, and other systematics, each successive exposure is rotated by 90 degrees, and a small ($\sim 0.05 \arcsec$) dither pattern is applied. We reduce the raw data following the procedure detailed in \citet{Richard_2021}. Details of the observations for both clusters are summarised in Table\,\ref{tab:data_MUSE}.

\subsection{X-ray data}

We searched the \textit{XMM-Newton} archive for publicly available observations of the two systems of interest. MACS\,J0242 was observed for a total of 70\,ks (OBSID:0673830101), and MACS\,J0949 for a total of 36\,ks (OBSID:0827340901). We analysed the two observations using \textsc{XMMSAS v17.0}, and the most up-to-date calibration files. We used the XMMSAS tools \texttt{mos-filter} and \texttt{pn-filter} to extract light curves of the observations and filter out periods of enhanced background, induced by soft proton flares. After flare filtering, the available clean exposure time is 61\,ks (MOS) and 53\,ks (PN) for MACS\,J0242, and 35\,ks (MOS) and 34\,ks (PN) for MACS\,J0949.

\section{Spectroscopic \& Photometric Analyses}
\label{sec:Analysis}

In this section, we present the key steps to obtain cluster galaxy catalogues and (candidate) background multiple image systems for both MACS\,J0242 and MACS\,J0949: from the source extraction to the selections of galaxies and identification of cluster galaxies specifically, using both the multi-band imaging in hand for the two clusters as well as the spectroscopy from VLT/MUSE.

\subsection{Spectroscopic analysis} \label{subsec:spectrcopic_catalogues}

We here present the analysis of the spectroscopic observations described in Sect.\,\ref{sec:Data_spectro}.
In spite of the field of view of the MUSE cubes, $1\arcmin \times 1\arcmin$, being smaller than that of \textit{HST} or DES, we can still access the redshift of a large number of foreground, cluster and background galaxies.

In order to detect specifically multiple image systems, we use \texttt{MUSELET} (MUSE Line Emission Tracker), a package of \texttt{MPDAF} (Muse Python Data Analysis Framework) which removes the constant emission from bright galaxies in the field, and is optimised for the detection of the faintest objects. For more details about the technique, we refer the reader to \citep{2016ascl.soft11003B} and \citep{piqueras2017mpdaf}.
We go through each of the 3681 slices of this subtracted MUSE datacube, and identify the bright detections.

We complete this technique with \texttt{CatalogueBuilder} \citep[see][]{Richard_2021} for a thorough and systematic analysis. The latter embeds the \texttt{MUSELET} analysis, but also uses a modified version of \texttt{MARZ} \citep[see][]{Hinton_2016}, which is better tuned to the resolution and spectral profiles specific to MUSE data. \texttt{CatalogueBuilder} also uses the position data of the deepest field available (in this case \textit{HST}/ACS).
These make it easier to confirm the likely source of the multiple images which we are looking for.
Using the spectroscopic information, we adjust with our own custom \textit{redshifting} routine the detected spectra to the known absorption lines, and notably [OII], [OIII] and Ly-$\alpha$. We then obtain catalogues containing coordinates and redshifts, such as Tables \ref{tab:spectro_mul_m0242} and \ref{tab:spectro_mul_m0242_additional}. We also consider multiple detections within a radius of $< 0.5\arcsec$ and a redshift separation of $\delta z < 0.05$ to be a unique object. 
All redshifts are supposed known with a precision estimated to $\delta z = 0.0001$.

We can associate to these detections Signal-to-Noise (S/N) ratios. As we also know the type of pattern the absorption lines should match, we can use the S/N ratio and spectral patterns to define different confidence levels. We only keep in all catalogues, including for example in Sect.\,\ref{sec:spectroscopic_detections_of_interest}, detections judged to be “good” or “excellent” (identifiers 3 and 4 in \texttt{MARZ} and \texttt{CatalogueBuilder}).
In the case of several detections representing a same object, we merge them keeping the best quality of detection.

\begin{figure*}
\begin{minipage}{0.48\textwidth}
\centerline{\includegraphics[width=1\textwidth]{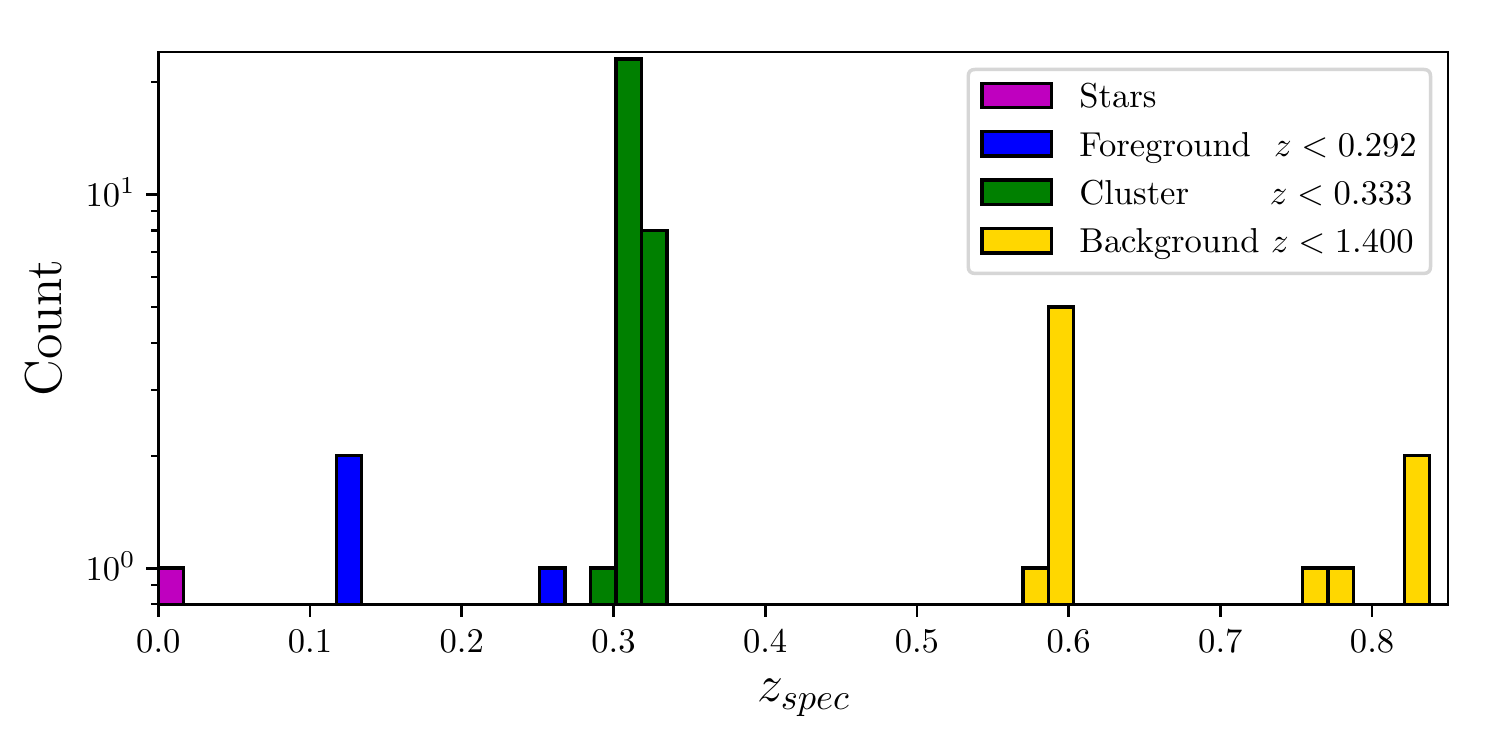}}
\end{minipage}
\hfill
\begin{minipage}{0.48\linewidth}
\centerline{\includegraphics[width=1\textwidth]{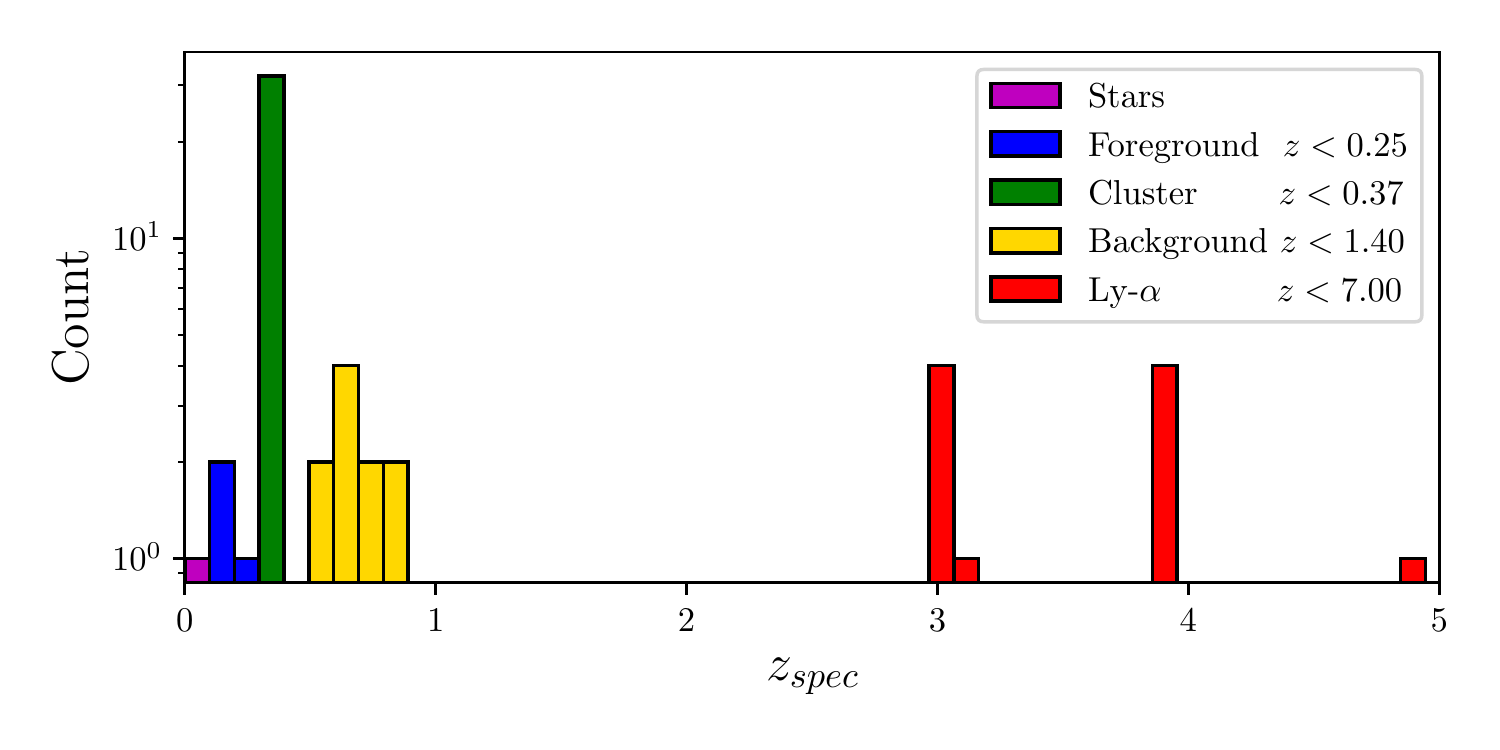}}
\end{minipage}
\begin{minipage}{0.48\textwidth}
\centerline{\includegraphics[width=1\textwidth]{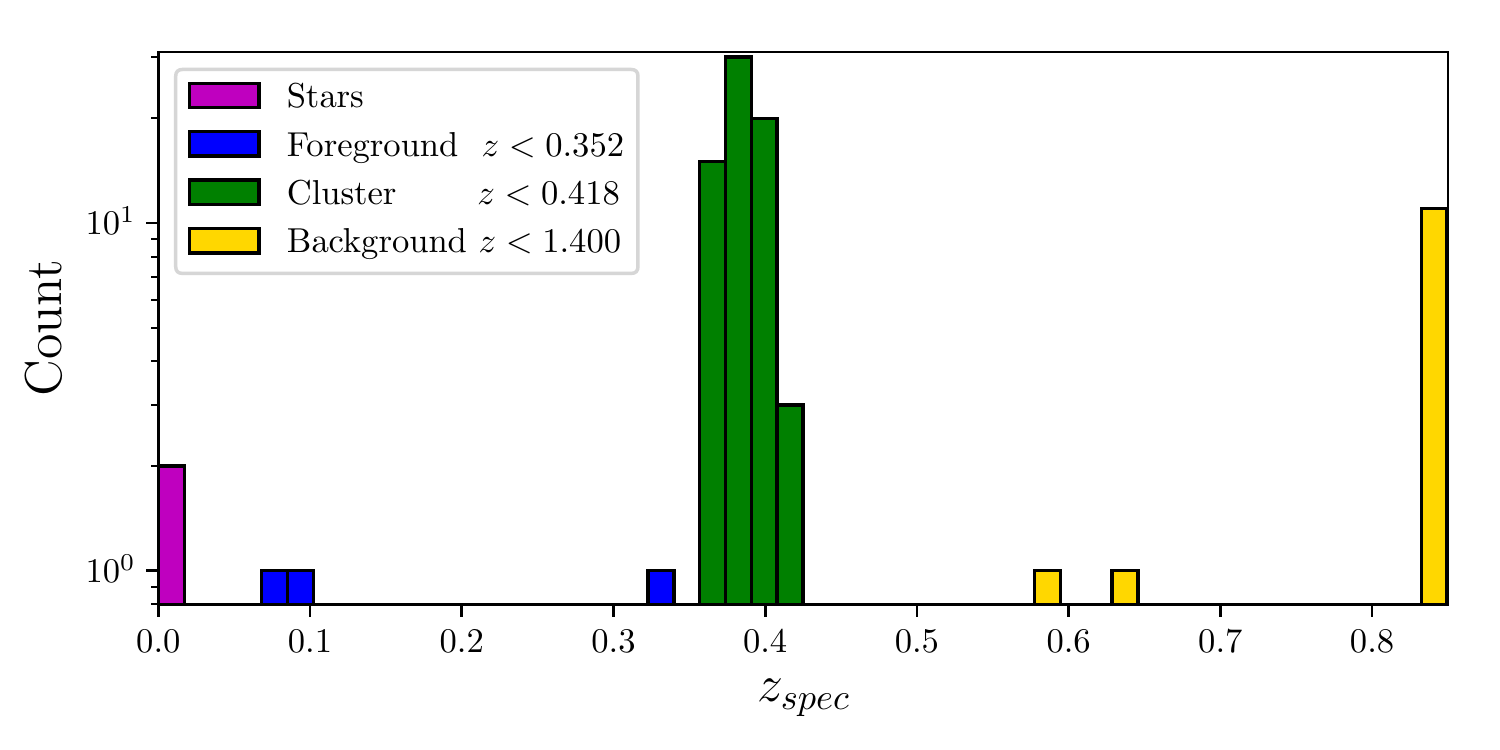}}
\end{minipage}
\hfill
\begin{minipage}{0.48\linewidth}
\centerline{\includegraphics[width=1\textwidth]{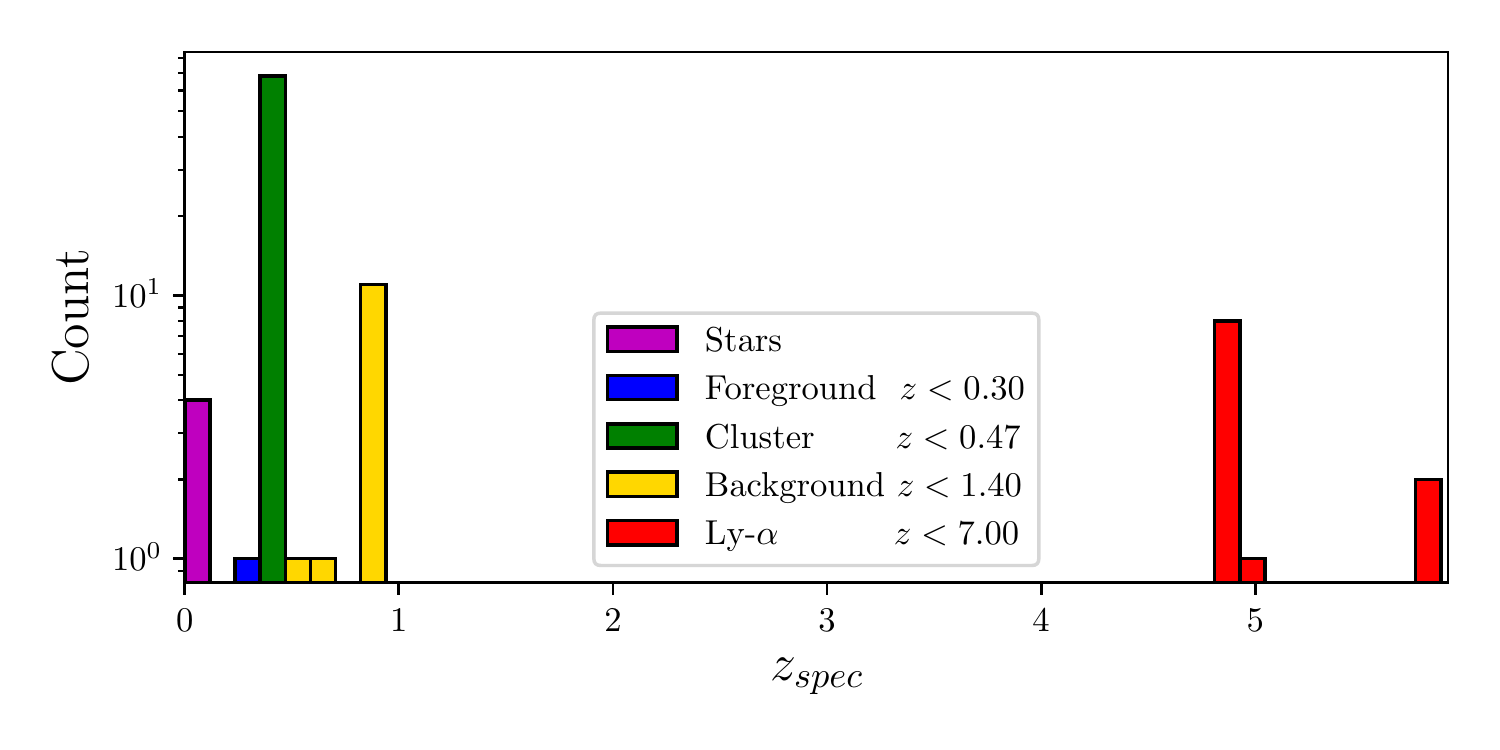}}
\end{minipage}
\caption{Redshift distribution of all MUSE detected objects. \textit{Top row}: Cluster MACS\,J0242. Objects identified as being in the cluster are shown in green, while foreground and background objects are shown in blue and yellow respectively. We highlight Lyman-$\alpha$ emitters in red. At last, objects within the Milky Way (stars, etc.) are displayed in purple. \textit{Left panel:} Redshift distribution of objects located at small redshifts $z < 1$. -- \textit{Right panel:} Redshift distribution of all objects with a measured redshift.
\textit{Bottom row}: Cluster MACS\,J0949. \textit{Left panel:} Redshift distribution of objects located at small redshifts $z < 1$. -- \textit{Right panel:} Redshift distribution of all objects with a measured redshift.}
\label{fig:spectro_histo}
\end{figure*}

The distribution of redshifts in each cluster is shown in Fig.\,\ref{fig:spectro_histo} for the full MUSE frame. We measure 36 and 96 good spectroscopic redshifts in MACS\,J0242 and MACS\,J0949 respectively. Due to the small statistics, this distribution is not Gaussian but it is sufficient to constrain the redshift of the clusters, which we estimate to be $0.300 \leq z \leq 0.325$ and $0.36 \leq z \leq 0.41$ for MACS\,J0242 and MACS\,J0949 respectively. For the current analysis, we define the redshift of each cluster by that of their BCG, i.e. respectively $0.3131$ and $0.383$ for MACS\,J0242 and MACS\,J0949 respectively.

\subsection{Photometric analysis}
\label{sec:photoanalysis}
\subsubsection{Source extraction}
\label{sec:Extraction}

We first align all images from a given instrument (\textit{HST}/ACS, \textit{HST}/WFC3, \textit{HST}/WFPC2 and DESI) to the same $wcs$ coordinates, and pixelate them accordingly to allow for direct colour comparison of detected objects.
In order to extract all detected objects from the multi-band imaging in hand for each cluster, we run the \texttt{SExtractor} software \citep{Sextractor_ref} in dual-image mode, for each pass-band of each instrument. For each instrument, we adopt a reference pass-band and a position of reference. The former sets the Kron-like magnitude of each detection, while the latter sets its location. 
The number of bands per instrument as well as the reference pass-band used are listed in Tables \ref{tab:global_Sextractor_remaining galaxies_m0242} and \ref{tab:global_Sextractor_remaining galaxies_m0949} for MACS\,J0242 and MACS\,J0949 respectively.

For each instrument, we then apply several cuts and selection criteria to the output catalogues from \textsc{SExtractor}. That allows us to build a complete multi-band catalogue composed only of galaxies.
We summarise the different steps of this process:
    
    (i) All detections without reliable magnitude measurements (i.e. MAG$\_$AUTO$=$-99) and incomplete (or corrupted) data are removed from all catalogues. This includes isophotal data and memory overflow that occurs during deblending or extraction.
    
    (ii) All objects with a stellarity greater than $0.2$ are removed as they are likely to be stars rather than galaxies. We additionally mask all detections very close to bright stars.
    
    (iii) For a given cluster, only objects detected in all pass-bands are kept.
    
    (iv) All objects with a Signal-to-Noise ratio (S/N) smaller than 10 are removed.

Tables \ref{tab:global_Sextractor_remaining galaxies_m0242} and \ref{tab:global_Sextractor_remaining galaxies_m0949} are listing the number of detections remaining once each of these criteria are applied for each instrument, for MACS\,J0242 and MACS\,J0949 respectively.

\begin{table}
	\centering
	\caption{Number of detections (Nod) after each source extraction selections as listed in Sect.\,\ref{sec:Extraction} for MACS\,J0242.}
	\label{tab:global_Sextractor_remaining galaxies_m0242}
	\begin{tabular}{lccc}
	    \hline
		\hline
		Observable & DES & \textit{HST}/WFPC2 & \textit{HST}/ACS\\
		\hline
		Number of bands & 3 & 2 & 1\\
		Reference band & z & F814W  & F606W \\
		\hline
		Nod (0) & 186 & 808 & 559\\
		Nod (i) & 185 & 540 & 559\\
		Nod (ii) & 180 & 492 & 456\\
		Nod (iii) & 180 & 429 & 456\\
		Nod (iv) & 142 & 202 & 402\\
		Colour-Magnitude & 51 & 45 & 179\\
		\hline
		Final & & 58 & \\
		\hline
		\hline
	\end{tabular}
\end{table}

\begin{table}
	\caption{Number of detections (Nod) after each source extraction selections as listed in Sect.\,\ref{sec:Extraction} for MACS\,J0949.}
	\label{tab:global_Sextractor_remaining galaxies_m0949}
\makebox[.48\textwidth][c]{
    \begin{tabular}{lcc}
\hline
\hline
Observable & \multicolumn{1}{c}{\textit{HST}/WFC3} & \multicolumn{1}{c}{\textit{HST}/ACS} \\
		\hline
		Number of bands & 4 & 2\\
		Reference band & F160W & F814W \\
		\hline
		Nod (0) & 3114 & 3055 \\
		Nod (i) & 2388 & 2700 \\
		Nod (ii) & 2172 & 2639 \\
		Nod (iii) & 1648 & 2490 \\
		Nod (iv) & 773 & 1708 \\
		Colour-Magnitude & 42 & 172 \\
		\hline
		Final & \multicolumn{2}{c}{170} \\
		\hline
		\hline
\end{tabular}
}
\end{table}

\subsubsection{Spectroscopic redshift identification}

Now that we have a galaxy catalogue for each instrument, we can match our detection with spectroscopic redshift measurements from VLT/MUSE.
In order to ensure a MUSE detection corresponds to a photometric one, we compare the positions measured by \texttt{Sextractor} in the different filters for all objects, using a Haversine function\footnote{The Haversine angle reads as 
$$
    \begin{gathered}
    \begin{aligned}
    \mathcal{H} = 2 \arcsin \sqrt{ \sin^2 \left( \frac{\delta_2 - \delta_1}{2} \right) +  \cos \delta_1 \cos\delta_2 \sin^2 \left( \frac{\alpha_2 - \alpha_1}{2}\right) }.
    \end{aligned}
    \end{gathered}
    \label{eq:haversine}
$$}.
If the separation angle between objects from the spectroscopic and the photometric catalogues is smaller than $0.5\arcsec$, we consider the detection to be of the same objects, and hence associate the spectroscopic redshift to the photometric detection. This error is equal to 2.5 MUSE pixels, and captures the positional uncertainty on spectroscopic detections.

Out of this step, we attribute a spectroscopic redshift to 20, 25, and 25 sources in the DES, \textit{HST}/WFPC2 and \textit{HST}/ACS catalogues for MACS\,J0242. In the case of MACS\,J0949, we attribute a spectroscopic redshift to 54, and 49 sources in the \textit{HST}/ACS and \textit{HST}/WFC3 catalogues.

\subsubsection{Cluster galaxy selection} \label{sec:Photo_selection}
The next step is the identification of cluster galaxies specifically. For that we are using colour-magnitude selections for each clusters.

The first step consists in applying the red sequence technique \citep[e.g.][]{Gladders_2000}. Using the catalogues after source extraction selections and spectroscopic redshift identification, we compute for both clusters a series of colour-magnitude (CM) diagrams.
We compute these for each instrument. As each pass-band represents a magnitude, we can respectively compute 3 and 1 CM diagrammes for DES and \textit{HST}/WFPC2 for MACS\,J0242 (none for \textit{HST}/ACS as only one band is available), and 1 and 6 for \textit{HST}/ACS and \textit{HST}/WFC3 for MACS\,J0949.

\begin{figure*}
\begin{minipage}{0.48\textwidth}
\centerline{\includegraphics[width=1\textwidth]{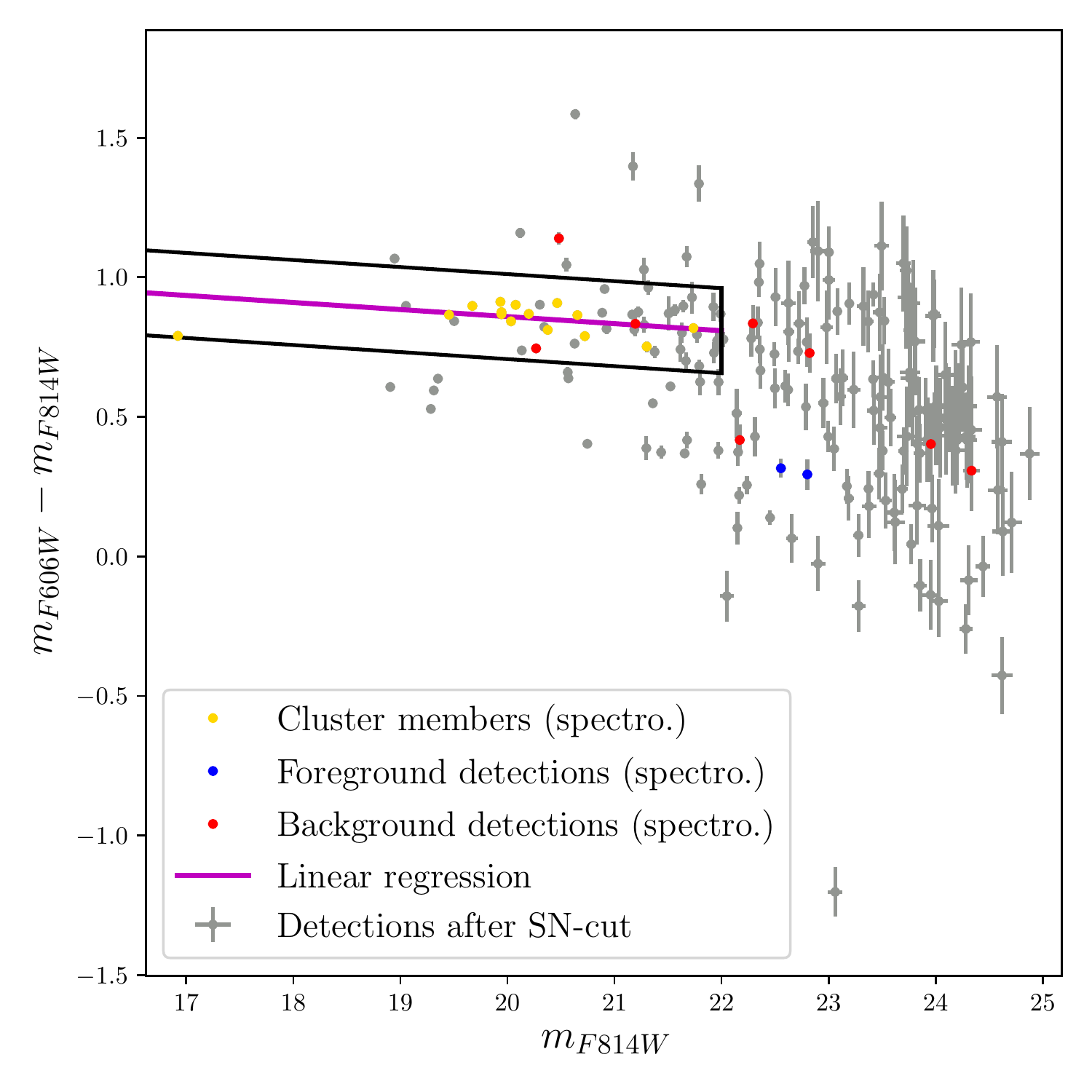}}
\end{minipage}
\hfill
\begin{minipage}{0.48\linewidth}
\centerline{\includegraphics[width=1\textwidth]{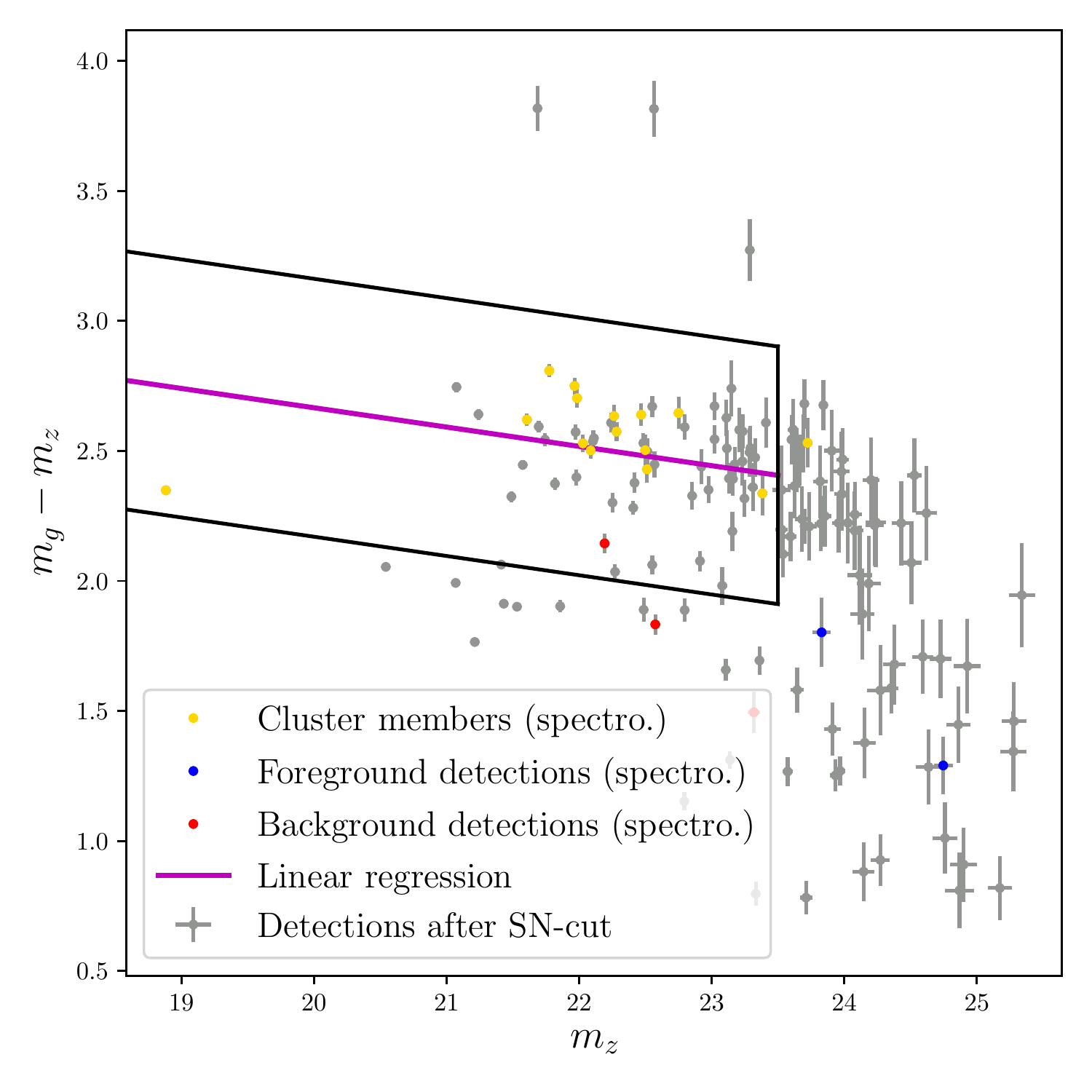}}
\end{minipage}
\begin{minipage}{0.48\textwidth}
\centerline{\includegraphics[width=1\textwidth]{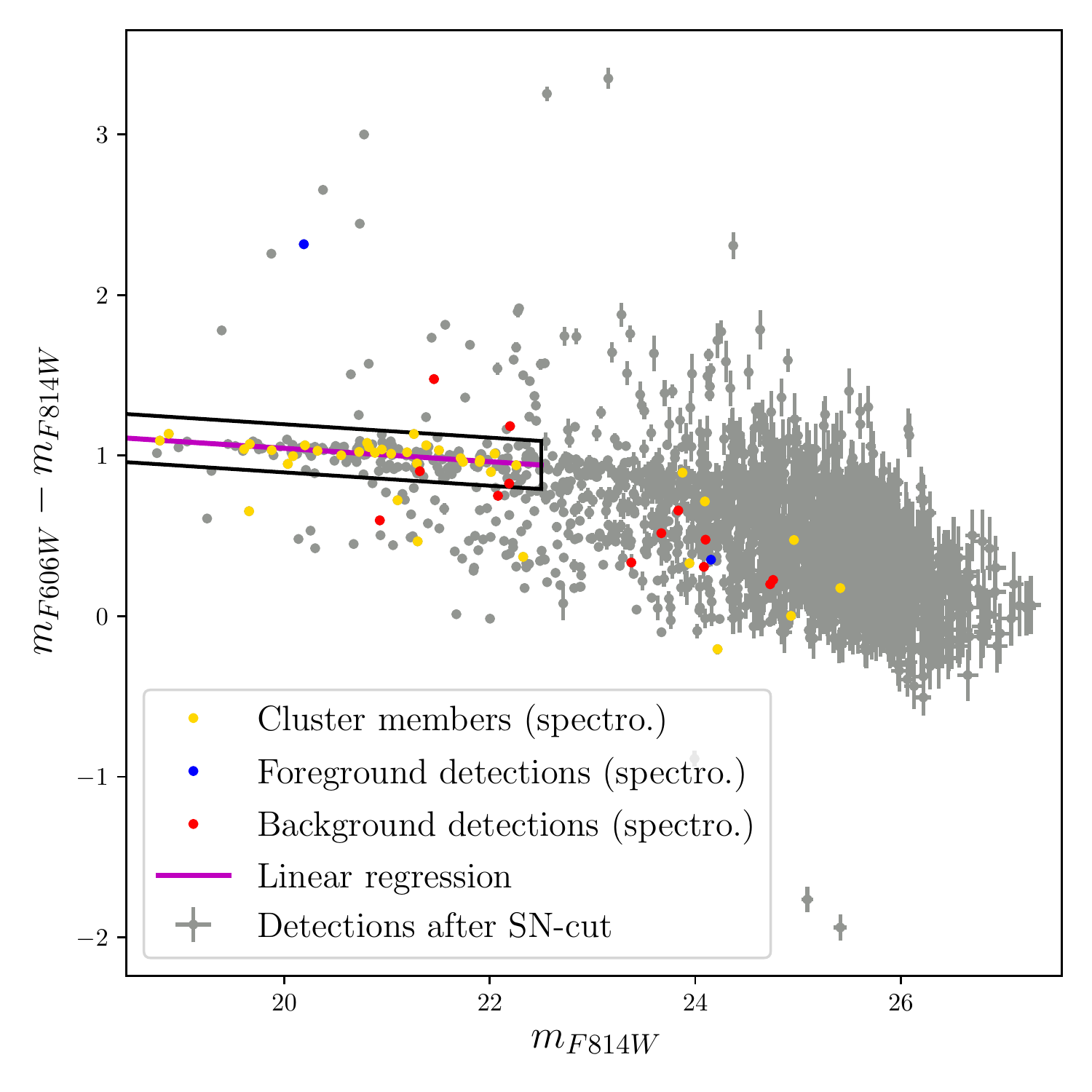}}
\end{minipage}
\hfill
\begin{minipage}{0.48\linewidth}
\centerline{\includegraphics[width=1\textwidth]{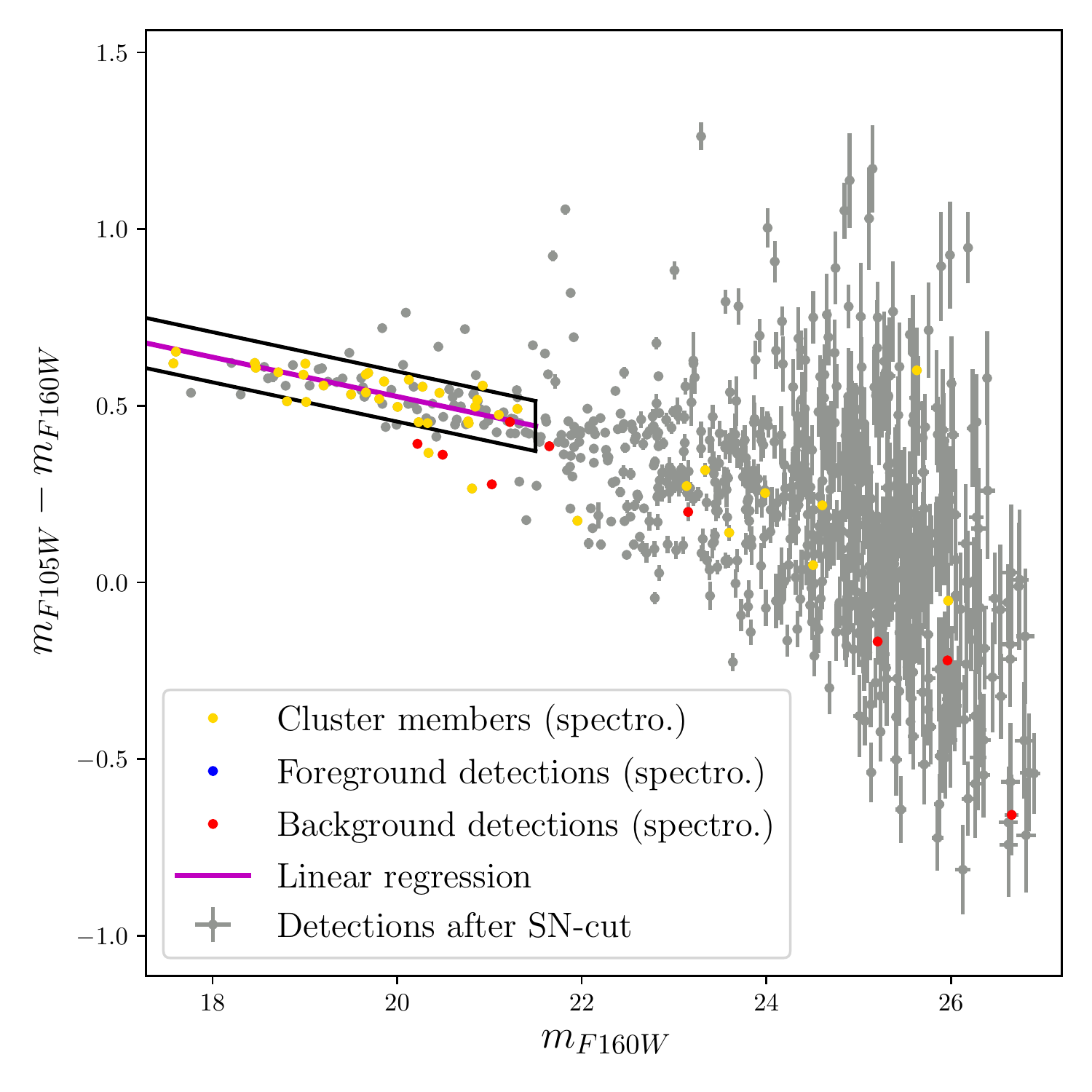}}
\end{minipage}
\caption{Colour-magnitude diagrams. \textit{Top row:} Cluster MACS\,J0242. \textit{Left panel:} Instrument \textit{HST}/WFPC2 -- $m_{\mathrm{F814W}}$ vs ($m_{\mathrm{F606W}} - _{\mathrm{F814W}}$). \textit{Right panel:} Instrument DES -- $m_{\mathrm{z}}$ vs ($m_{\mathrm{g}} - m_{\mathrm{z}}$). Grey filled circles (with their error bars) have successfully passed all selections described in Section \ref{sec:Extraction}. The magenta line represents the main sequence regression. Blue, gold and red dots represent spectroscopic detections of foreground, cluster and background objects respectively. \textit{Bottom row:} Cluster MACS\,J0949. \textit{Left panel:} Instrument \textit{HST}/ACS -- $m_{\mathrm{F814W}}$ vs ($m_{\mathrm{F606W}} - m_{\mathrm{F814W}}$). \textit{Right panel:} Instrument \textit{HST}/WFC3 -- $m_{\mathrm{F160W}}$ vs ($m_{\mathrm{F105W}} - m_{\mathrm{F160W}}$).}
\label{fig:CM_diag}
\end{figure*}

As shown in Fig.\,\ref{fig:CM_diag}, cluster members are expected to follow a main sequence (magenta line).
To calibrate our selections, we use spectroscopically confirmed cluster members.
We then remove all detections with a magnitude exceeding $m_{\mathrm{max}}$, which varies depending on instruments and filters.
For MACS\,J0242, we have $m_{\mathrm{max}} = 22$ for \textit{HST}/WFPC2, 23.5 for DES/z, and 24.5 for DES/r. For MACS\,J0949, we have $m_{\mathrm{max}} = 21.5$ for \textit{HST}/WFC3 and 22.5 for \textit{HST}/ACS.
We then perform a linear regression and obtain the main sequence. We give in Appendix\,\ref{sec:additonal_CM_CC_data} the fits for all colour-magnitudes used for both clusters.

Galaxies selected as cluster members are galaxies which have a colour within $2 \sigma_{C}$ of the main red sequence for \textit{HST}/ACS and \textit{HST}/WFC3, and within $3 \sigma_C$ for \textit{HST}/WFPC2 and DES. $\sigma_{C}$ is the weighed colour standard deviation of the spectroscopically confirmed cluster galaxy sample.
These limits are highlighted as black rectangles in Fig.\,\ref{fig:CM_diag}.
For an instrument with more than 2 pass-bands, we can compute more than one CM diagram, and thus only retain cluster member identifications compatible with \textit{all} colour-magnitude diagram selections.
We summarise in Tables\,\ref{tab:global_Sextractor_remaining galaxies_m0242} and \ref{tab:global_Sextractor_remaining galaxies_m0949} for MACS\,J0242 and MACS\,J0949 respectively, the number of galaxies identified as cluster members per instrument once these colour-magnitude selections are applied.
In some cases, spectroscopically confirmed cluster galaxies fall outside the colour-magnitude selection. These objects are ultimately conserved in our cluster galaxy catalogue. However, we do not include them in the CM cut counts, to show the effect of the photometric selection.

\subsubsection{Instrument catalogue combination}

We now assemble the galaxy catalogues for each instrument before merging these into a final cluster galaxy catalogue for each cluster. We match the coordinates of sources with the already defined $0.5\arcsec$ separation angle.

MACS\,J0242 and MACS\,J0949 were imaged with different instruments, and thus have different coverage.
We define the camera of reference as the camera with the highest resolution. In the case of the both clusters, it is \textit{HST}/ACS, but the reference band is chosen as F606W for MACS\,J0242, and F814W for MACS\,J0949.
MACS\,J0242 was observed with \textit{HST}/ACS in only one band.
Moreover, MACS\,J0242 was observed with \textit{HST}/WFPC2 in 2 pass-bands, but the shape of the camera field of view does not cover the entire ACS field of view.
MACS\,J0242 has DES observations in 3 pass-bands, covering a wide field of view. However the quality of these observations is lower than the ones we have from space.
We therefore require for a given cluster member selected galaxy in \textit{HST}/ACS to be at least present in DES or WFPC2 in order to be included into the final cluster member catalogue.
MACS\,J0949 was imaged with \textit{HST}/ACS and WFC3 cameras.
\textit{HST}/WFC3 has a smaller field of view than ACS. 
We detected multiply imaged systems out of the WFC3 field of view. In order to account for the gravitational effect of individual galaxies on these systems, we include all galaxies detections from at least one camera to our galaxies catalogue.

Finally, cluster galaxies located at a distance larger than 40\arcsec from the cluster centre and with a magnitude difference to the BCG of $\Delta m > 4$ are ignored. Due to their small mass, these galaxies would only have a very small impact on the strong lensing configurations observed.

\subsection{Final catalogues}

\subsubsection{Cluster galaxy catalogues}

Sect.\,\ref{sec:photoanalysis} describes all the steps for the identification of cluster members, including colour-magnitude selections as well as spectroscopic identifications.
All galaxies identified as cluster members and used for our lensing modelling are listed in Appendix, in Tables\,\ref{tab:cl_members_m0242} and \ref{tab:cl_members_m0949} for MACS\,J0242 and MACS\,J0949. Our final catalogues include 58 and 170 galaxies for MACS\,J0242 and MACS\,J0949 respectively.

In order to probe the robustness of our catalogues, we conducted the following verification analysis. We isolated only the spectroscopic detections, and then reinjected them into our photometric selection. We found respectively 15 out of 16 and 34 out of 34 galaxies retained within the photometric selection for MACS\,J0242 and MACS\,J0949.
As these spectroscopic detections were used to define these selections, they are expected to be selected. Thus, in order to estimate the contamination by galaxies out of the cluster redshift boundaries, we examined the number of selected spectroscopic detections out of the cluster.
We find a maximum 2 (2) out of 54 (97) galaxies of our sample contaminants, i.e. $4\%$ ($2\%$) contamination of our sample in cluster MACS\,J0242 (MACS\,J0949).
Thus we are confident in our galaxy selection.
Nevertheless, for accuracy, we removed these known out-of-cluster galaxies from the final catalogue.

\subsubsection{Multiple image systems}

In Sect.\,\ref{subsec:spectrcopic_catalogues}, we described the preliminary steps leading to the multiple image system catalogue. At this point, this is simply a catalogue of reliable detections with redshift $z > 0.6$.
The second step in the identification of multiple image systems is to look for similarities between these detections, starting with their spectra. We then look at their positions and see if they are compatible with a lensing geometry.
The MUSE field of view being narrower than the \textit{HST} one, one can also look at the colour and morphology of possible multiple images.
If a given set of multiple images presents at the same time compatible positions, colours, morphologies and, if available, redshift, we consider them as a multiple image system.

\begin{figure*}
\begin{minipage}{0.48\textwidth}
\centerline{\includegraphics[width=1\textwidth]{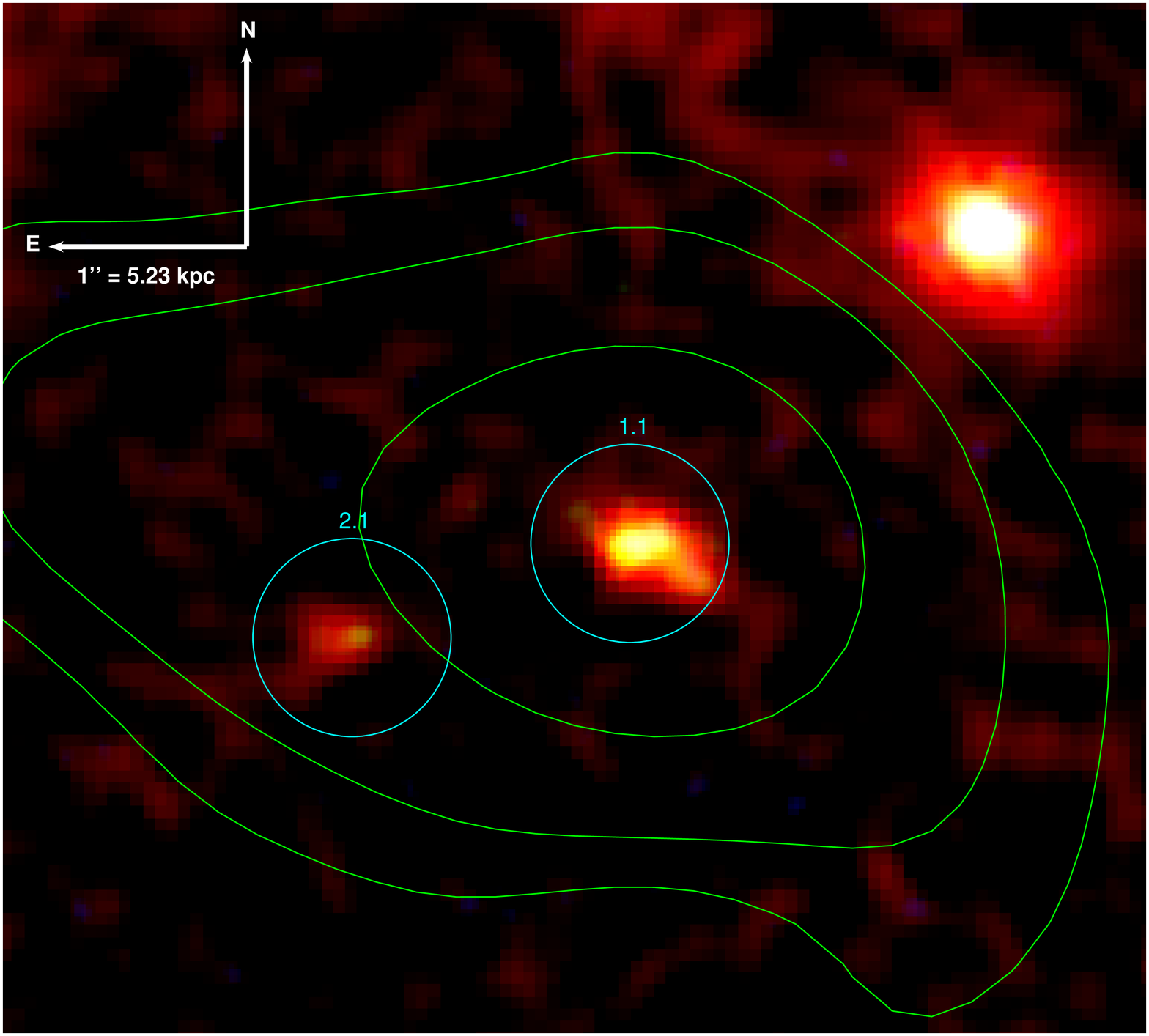}}
\end{minipage}
\hfill
\begin{minipage}{0.48\linewidth}
\centerline{\includegraphics[width=1\textwidth]{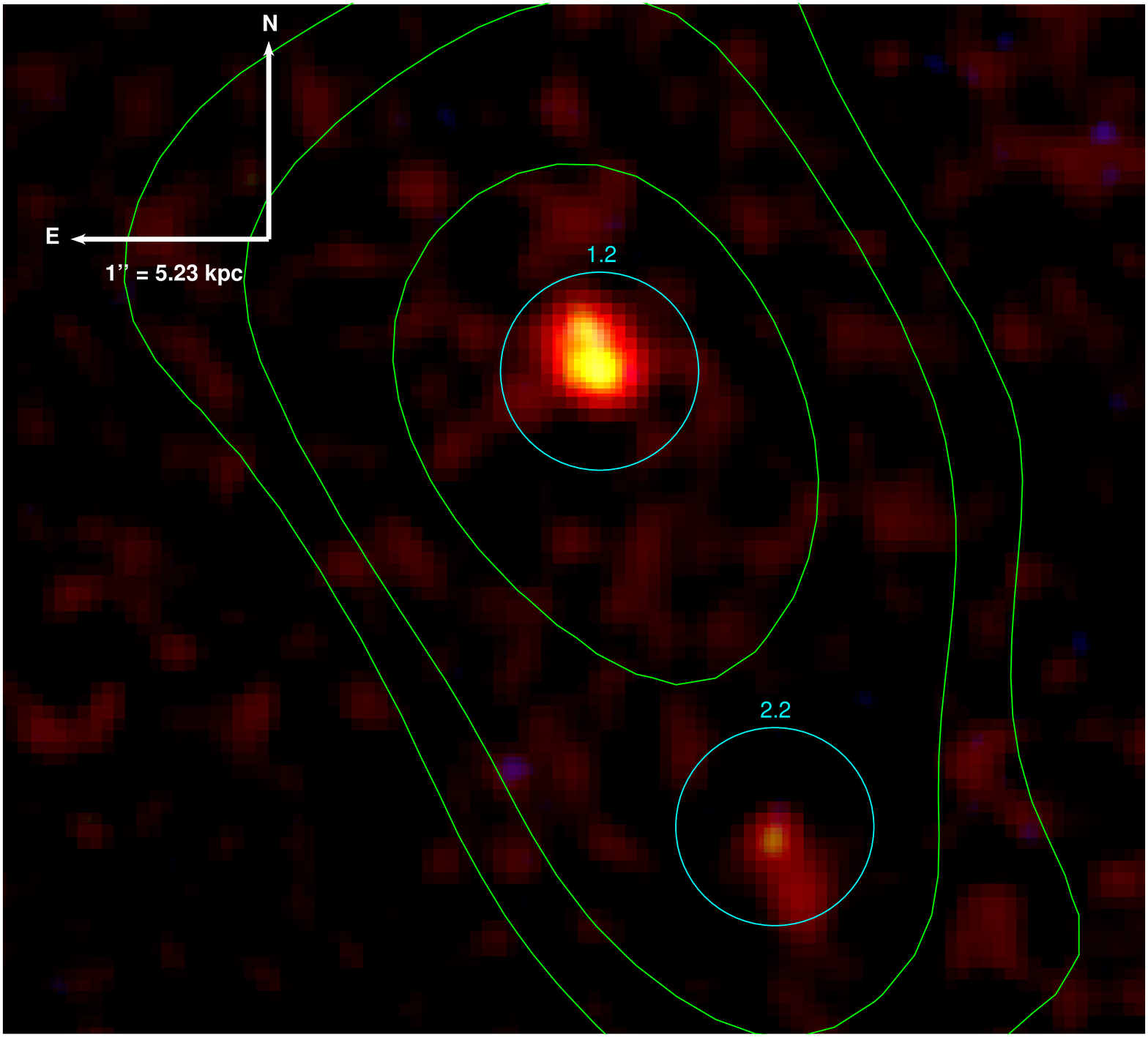}}
\end{minipage}
\begin{minipage}{0.48\textwidth}
\centerline{\includegraphics[width=1\textwidth]{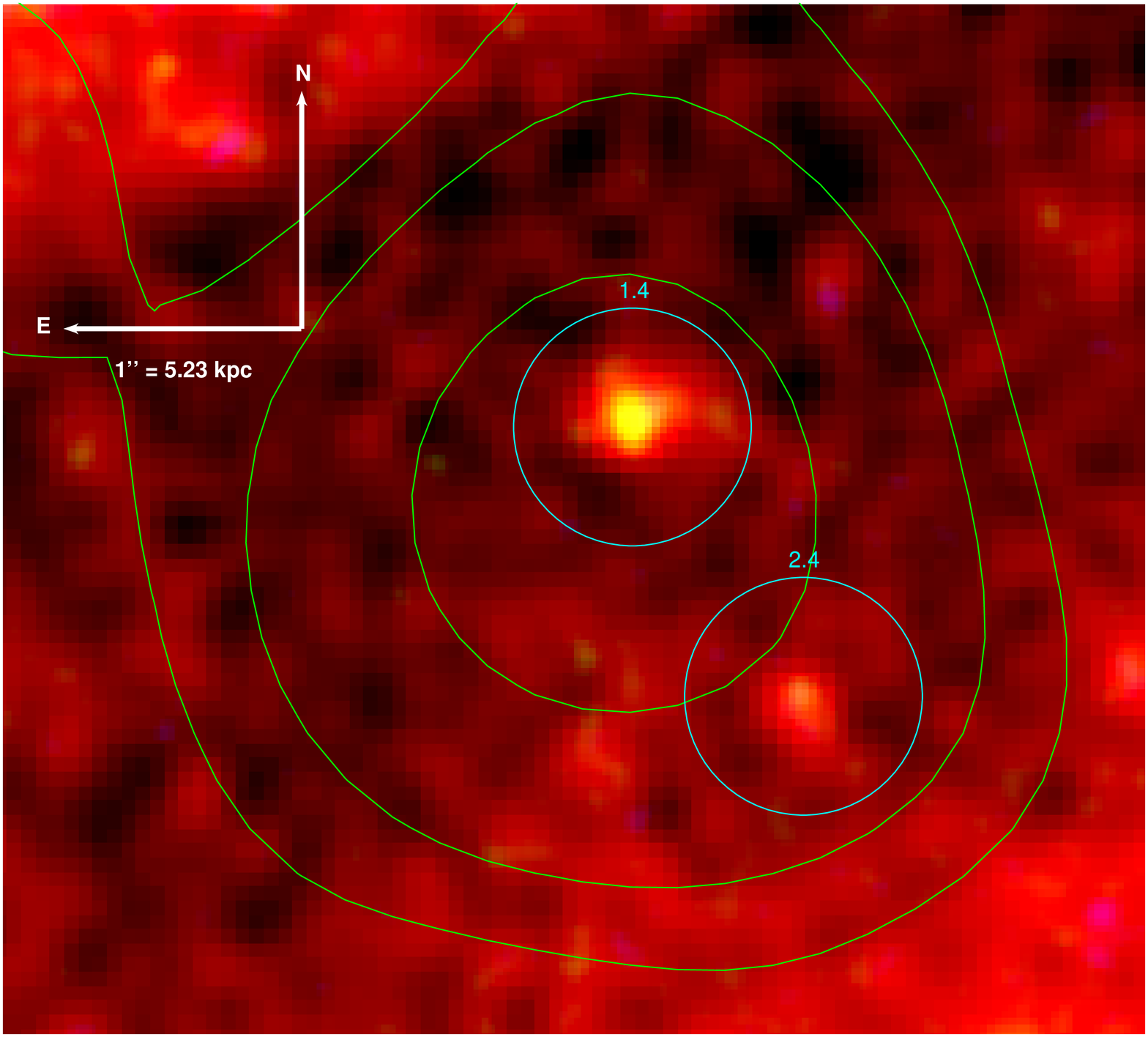}}
\end{minipage}
\hfill
\begin{minipage}{0.48\linewidth}
\centerline{\includegraphics[width=1\textwidth]{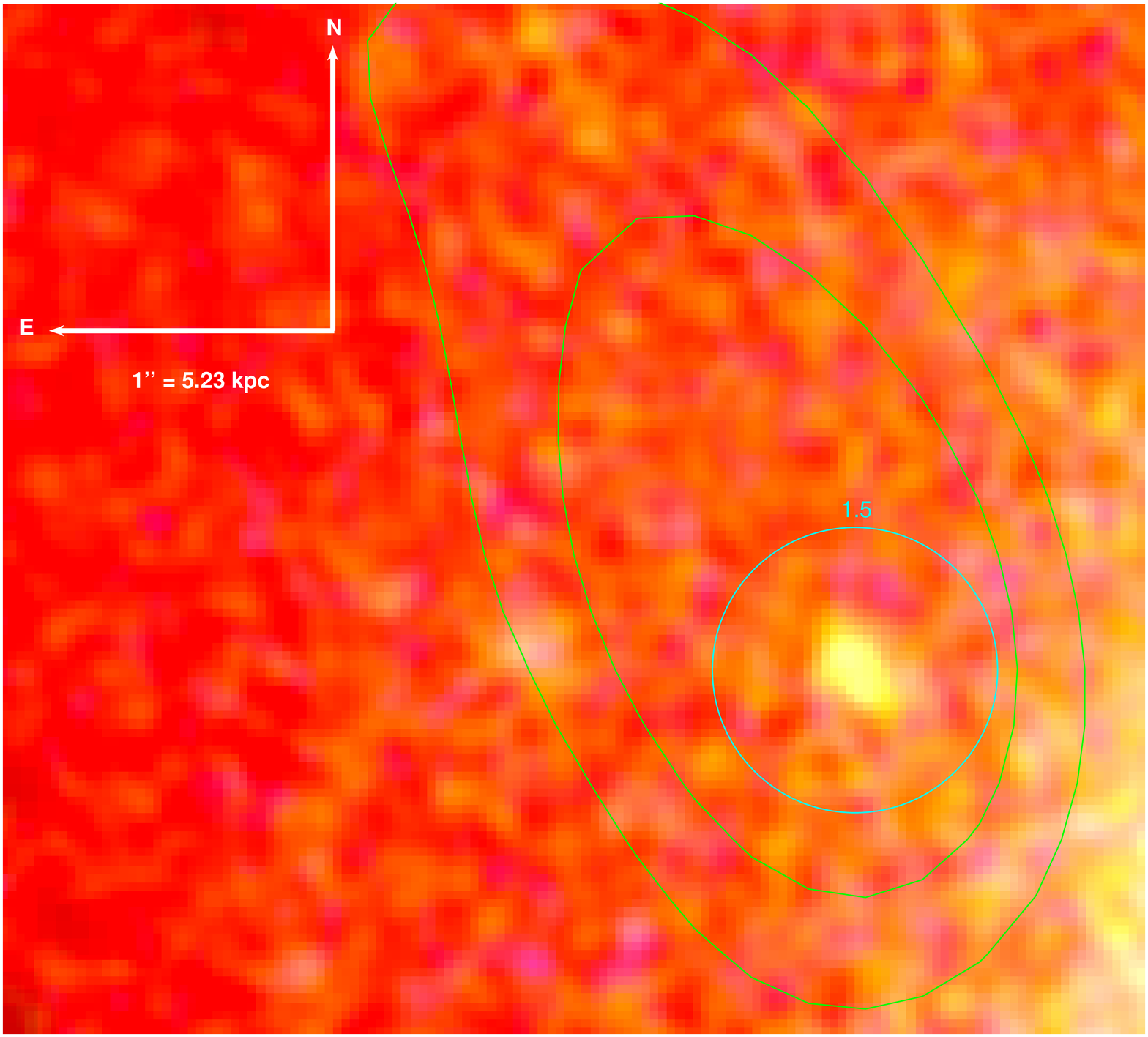}}
\end{minipage}
\caption{\textit{HST} composite colour image of the four multiple images of System\,1 detected in MACS\,J0949 with VLT/MUSE observations. Colours were enhanced to outline the multiple images. Labelled cyan circles show the positions of the multiple images and correspond to the peak of the Lyman-$\alpha$ emission. The green contours show flux density levels at $1.500$, $2.125$ and $4.000 \times 10^{-20}$ erg\,s$^{-1}$ cm$^{-2}$\,\AA$^{-1}$.}
\label{fig:m0949_spectro_multimages}
\end{figure*}

In Fig.\,\ref{fig:m0949_spectro_multimages}, we show a colour composite \textit{HST} image of four MUSE detections, 4 multiple images of the same galaxy located at redshift $z = 4.89$. 
In the case of MACS\,J0949, we force extract emission from the MUSE cube corresponding to the location of multiple images previously identified by the RELICS collaboration (obtained through private communication); we only reveal marginal identification as explained in Sec.\,\ref{subsubsec:mass_model_m0949}. The final list of system used in this analysis is presented in Table\,\ref{tab:spectro_mul_m0949}.

\section{Strong Lensing Mass modelling}
\label{sec:Mass-modelling}

The mass distribution of each cluster is reconstructed using the \texttt{Lenstool} software\footnote{\url{https://projets.lam.fr/projects/lenstool/wiki}} \citep{1996ApJ...471..643K, 2007NJPh....9..447J}, in its parametric mode. 
The optimisation is performed in the image plane with a Markhov Chain Monte-Carlo algorithm (MCMC) assuring the sampling of parameter space. It optimises the predicted positions of multiple images while fitting an underlying mass distribution composed of large-scale halo(s) to describe the overall cluster potential, and small-scale halos to account for local perturbers such as cluster galaxies.

For both clusters, we describe any potential using a dual Pseudo-Isothermal Elliptical matter distribution \citep[dPIE, see][]{1993ApJ...417..450K} which, as described in \citet{eliasdottir2007matter}, has two different pivot scales: a \textit{core radius}, which describes the potential evolution due to the baryonic matter content, and a \textit{cut radius} that describes the dark matter potential.
A dPIE potential is described by seven parameters (excluding the redshift):
the central coordinates, the ellipticity $e$, the position angle $\theta$, the core and cut radii, $r_{\mathrm{core}}$ and $r_{\mathrm{cut}}$ respectively, and a fiducial central velocity dispersion $\sigma$.
The fiducial central velocity dispersion in \texttt{Lenstool} $\sigma$ relates to the true three dimensional central velocity dispersion with $\sigma_0 = \sqrt{3/2} \sigma$, as detailed in \citet{2019A&A...631A.130B}, Appendix C.

For each cluster, we assume one single large-scale dark matter halo to describe the overall cluster potential. It is described by a large velocity dispersion ($\sim 10^3$\,km.s$^{-1}$), a large core radius ($\sim 10^2$\,kpc) and large cut radius.
We optimise all the parameters of the potential, excluding the cut radius which we fixed to values $\geq$1\,Mpc as it is located far from the strong lensing region and thus cannot be constrained by multiple images only.
The position of each cluster halo is allowed to vary within 10\arcsec~of the cluster centre, i.e. the position of the BCG. The ellipticity of the halo is limited to values $<0.8$.
The cut radius is fixed to 1.5\,Mpc for both MACS\,J0242 and MACS\,J0949, as our investigation to model the ICM through lensing shows that this value provides a better fit to the X-ray observations (see our companion paper Allingham et al. in prep.).
This value is in agreement with \citet[][]{2018ApJ...864...83C}, taking in consideration the higher mass range of the clusters we are exploring here.

The BCG of each cluster is also modelled independently, using a dPIE potential. The BCG has a strong gravitational influence in the cluster core, and will thus impact the geometry of multiple images quite strongly \citep{NewmanAB_2013}. We fix their $r_{\mathrm{core}}$ to a small value of 0.30\,kpc for cluster MACS\,J0242 and 0.25\,kpc for MACS\,J0949. For their positions, position angle, and ellipticity, we fix their values to the shape parameters in outputs of \texttt{SExtractor}.
Finally, we only optimise its their velocity dispersion and cut radius.

Each individual cluster member is modelled by its own dPIE potential. Their positions, ellipticities and position angles are obtained with the photometric extraction.

We again assume a small but non-null value for $r_{\mathrm{core}}$. Their cut radii and velocity dispersions are optimised using their magnitude and assuming the Faber-Jackson scaling relation \citep{1976ApJ...204..668F}. All cluster members cut radii and velocity dispersions are rescaled with regard to a unique set of parameters ($r_{\mathrm{cut}, 0}$, $\sigma_{0}$).
This allows us to optimise each cluster galaxy potential using a remarkably small number of parameters.
$r_{\rm cut}$ and $\sigma$ are allowed to vary between 1 and 50\,kpc, and 100 and 300\,km.s$^{-1}$ respectively.
As mentioned earlier, the Faber-Jackson relation being scaled to a reference magnitude $mag_{0}$, we use the reference pass-band of the main camera for each cluster, ACS/F606W ($mag_{0}=20.0205$) and ACS/F814W ($mag_{0}=19.5085$) for MACS\,J0242 and MACS\,J0949 respectively.

As the centre of the cluster-scale halo and the BCG are aligned, the $r_{\mathrm{core}}$, $r_{\mathrm{cut}}$ and $\sigma$ parameters of both potentials are degenerate. 
Due to the limited number of lensing constraints, we proceed  incrementally to model the potential, to narrow the parameters space.
First, we include the BCG in the scaling relation of the cluster galaxies and optimise the cluster-scale halo and the scaling relation parameters as described above.
Second, we run a model with the BCG optimised independently, only optimising $r_{\rm cut}$ and $\sigma$ as explained above. However in this case, the cluster-scale halo parameters are allowed to vary within a restricted range, defined gaussianly around the best fit values obtained from the first model. 
This way, we can limit the degeneracy between the cluster-scale and BCG halos, and obtain physical values to describe the BCG potential.

\begin{figure}
    \centering
    \includegraphics[width=\columnwidth]{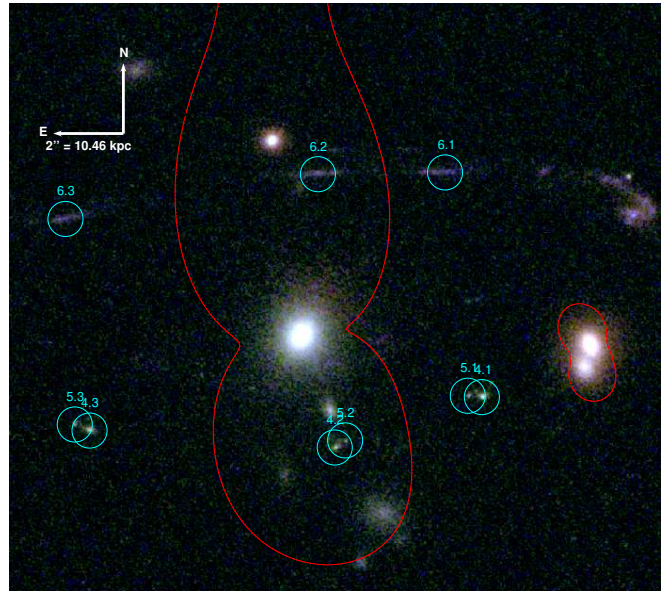}
    \caption{Composite colour \textit{HST} image of the Southern clump in MACS\,J0949.
    In cyan, we highlight the positions of the multiple images identified with \textit{HST}, and listed in Table\,\ref{tab:spectro_mul_m0949}. The external/tangential critical lines for a source at redshift $z = 3.65$ are represented in red -- this redshift being compatible with sources 4, 5 and 6, according to the best fit optimisation.}
    \label{fig:astro_south_m0949}
\end{figure}

Finally, we added a completely free dPIE potential south to the main cluster halo of MACS\,J0949. This structure has already been included in the public RELICS models and correspond to the location of three candidate multiply-imaged systems 4, 5 and 6 as shown in Fig. \ref{fig:astro_south_m0949}. We optimised their redshifts as well as the potential and to prevent nonphysically high value we imposed gaussian priors on $r_{\mathrm{core}}$, $r_{\mathrm{cut}}$ and velocity dispersion. 

\section{Results}
\label{sec:Results}

\subsection{Strong lensing mass models}

\subsubsection{MACS\,J0242 model}
\label{subsubsec:mass_model_m0242}

In MACS\,J0242, we detected six systems of multiple images with MUSE. Their positions and redshifts are given in Table\,\ref{tab:spectro_mul_m0242}.
We provide the best fit parameters of our model in Table\,\ref{tab:best_model}. The fixed values are highlighted by an asterisk. Our best-fit model yields predicted multiple images with a $rms$ of 0.39\arcsec of the observed positions.
The inclusion of an external shear component does not provide a significant improvement to the mass model, i.e. a $rms$ of 0.38\arcsec compared to our best-fit mass model of 0.39\arcsec.
This error is smaller than the positional error associated to spectroscopic detections. However, the error on the position of the multiply-lensed images is associated to their photometric detections, with much smaller positional error.

\begin{table}
	\centering
	\caption{List of multiple images detected with VLT/MUSE in MACS\,J0242. We here list their ID, coordinates, R.A. and Decl., given in degrees (J2000), and their measured spectroscopic redshift $z$.}
	\label{tab:spectro_mul_m0242}
	\begin{tabular}{lccc}
	    \hline
		\hline
		Id. & R.A. & Decl. & $z$\\
		\hline
		1.1 & 40.6574070  & -21.5383801  & 3.0627\\
		1.2 & 40.6575168  & -21.5387136  & 3.0627\\
        1.3 & 40.6531265  & -21.5473860  & 3.0627\\
        1.4 & 40.6446350  & -21.5392391  & 3.0627\\
        2.1 & 40.6453464  & -21.5336906  & 3.8681\\
        2.2 & 40.6411296  & -21.5407791  & 3.8681\\
        2.3 & 40.6419142  & -21.5436276  & 3.8681\\
        2.4 & 40.6546554  & -21.5416287  & 3.8681\\
        3.1 & 40.6580815  & -21.5363952  & 3.8682\\
        3.2 & 40.6454775  & -21.5404581  & 3.8682\\
        4.1 & 40.6523889  & -21.5446358  & 3.0615\\
        4.2 & 40.6499994  & -21.5316520  & 3.0615\\
        5.1 & 40.6529585  & -21.5386743  & 4.9492\\
        5.2 & 40.6432539  & -21.5482627  & 4.9492\\
        6.1 & 40.6499320  & -21.5354918  & 1.3010\\
        6.2 & 40.6541677  & -21.5382729  & 1.3010\\
        6.3 & 40.6463323  & -21.5366811  & 1.3010\\
        6.4 & 40.6479134  & -21.5470977  & 1.3010\\
		\hline
		\hline
	\end{tabular}
\end{table}

\begingroup
\renewcommand{\arraystretch}{1.2}
\begin{table*}
	\centering
	\caption{Best fit parameters of the strong lensing mass models for MACS\,J0242 and MACS\,J0949. We here list the central coordinates, $\Delta_{\alpha}$ and $\Delta_{\delta}$ in arcsec, relative to the centre, the ellipticity, $e$, the position angle in degrees, $\theta$, the core radius in kpc, $r_{\mathrm{core}}$, the cut radius in kpc, $r_{\mathrm{cut}}$, and the velocity dispersion in km.s$^{-1}$, $\sigma$, for each component of the model. The centres are taken to be respectively $(\alpha_c, \delta_c) = (40.649555, -21.540485)$\,deg and $(\alpha_c, \delta_c) = (147.4659012, 17.1195939)$\,deg for MACS\,J0242 and MACS\,J0949. The asterisks highlight parameters which are fixed during the optimisation.}
	\label{tab:best_model}
	\begin{tabular}{lccccccc}
	    \hline
		\hline
		 & $\Delta_{\alpha}$ & $\Delta_{\delta}$ & $e$ & $\theta$ & $r_{\textrm{core}}$ & $r_{\textrm{cut}}$ & $\sigma$\\
		\hline
		\hline
		& & & & MACS\,J0242 & & &\\
		\hline
        DM halo  & $-0.138_{-0.143}^{+0.085}$ & $0.136_{-0.179}^{+0.111}$ & $0.287_{-0.027}^{+0.037}$ & $17.884_{-1.830}^{+0.762}$ & $57.194_{-8.414}^{+6.044}$ & $1500^{\star}$ & $918.479_{-36.074}^{+28.984}$\\ 
        BCG & $0.044^{\star}$ & $-0.090^{\star}$ & $0.226^{\star}$ & $155.758_{-9.604}^{+10.766}$ & $0.300^{\star}$ & $177.575_{-57.950}^{+32.245}$ & $524.516_{-43.956}^{+58.810}$\\ 
        Galaxy catalogue &  &  &  &  & $0.030^{\star}$ & $5.625_{-1.808}^{+7.845}$ & $199.242_{-53.257}^{+30.721}$\\ 
		\hline
		\hline
		& & & & MACS\,J0949 & & &\\
		\hline
        DM halo & $-1.936_{-2.843}^{+0.215}$ & $-0.671_{-0.666}^{+0.565}$ & $0.249_{-0.045}^{+0.398}$ & $92.434_{-1.289}^{+0.570}$ & $116.246_{-51.661}^{+24.108}$ & $1500^{\star}$ & $1236.094_{-310.553}^{+59.307}$\\ 
        Southern halo & $4.800_{-0.464}^{+0.748}$ & $-60.133_{-1.417}^{+2.391}$ & $0.097_{-0.061}^{+0.294}$ & $128.629_{-27.521}^{+41.438}$ & $20.548_{-8.771}^{+31.596}$ & $232.502_{-119.902}^{+180.124}$ & $323.220_{-54.851}^{+120.202}$\\ 
        BCG & $0^{\star}$ & $0^{\star}$ & $0.475^{\star}$ & $120.130^{\star}$ & $0.250^{\star}$ & $98.044_{-34.342}^{+153.739}$ & $253.749_{-18.473}^{+196.474}$\\ 
        Galaxy catalogue & & & & & $0.150^{\star}$ & $23.135_{-2.053}^{+111.473}$ & $139.314_{-18.547}^{+25.804}$\\
		\hline
		\hline
	\end{tabular}
\end{table*}
\endgroup

\begin{figure*}
\begin{minipage}{0.48\textwidth}
\centerline{\includegraphics[width=1\textwidth]{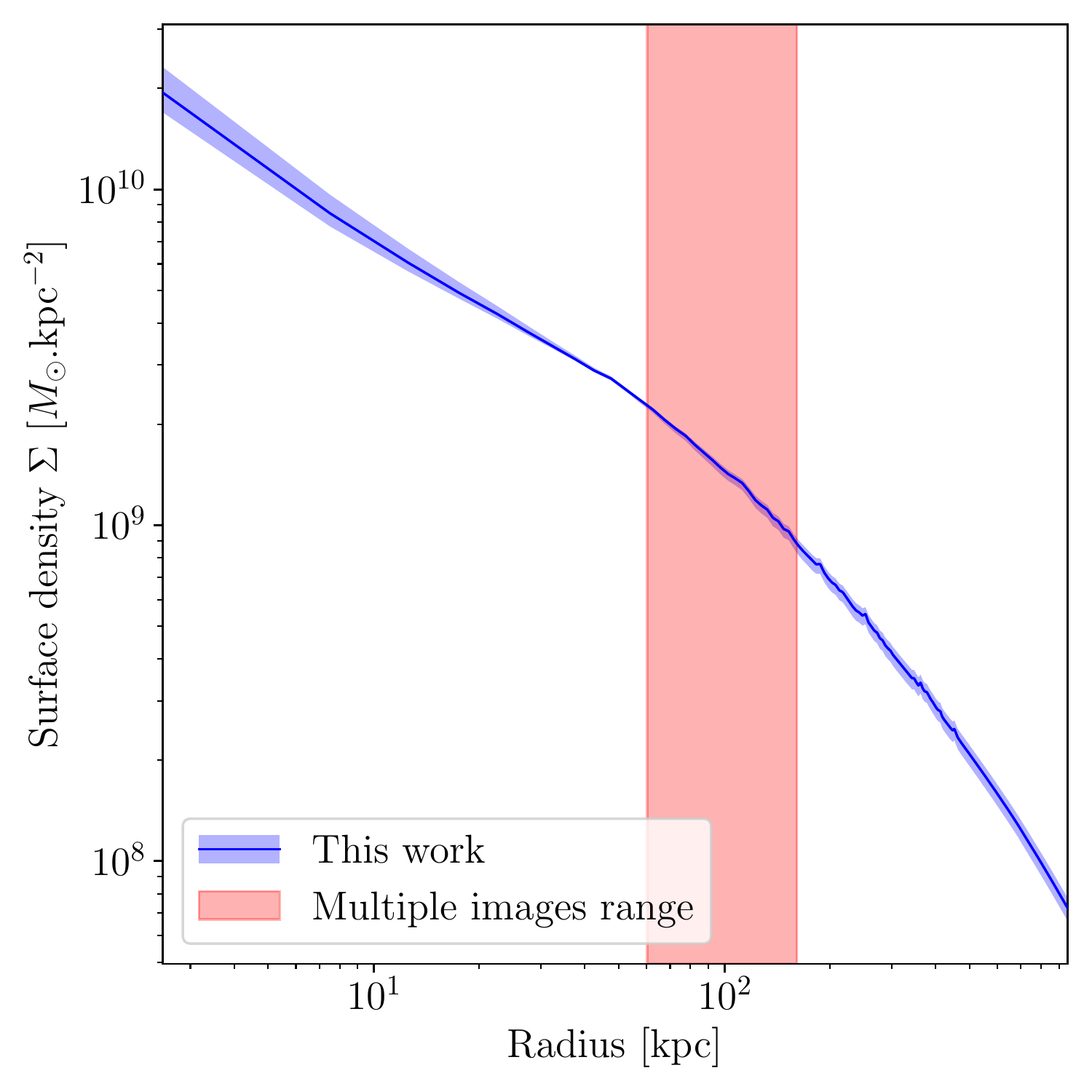}}
\end{minipage}
\hfill
\begin{minipage}{0.48\linewidth}
\centerline{\includegraphics[width=1\textwidth]{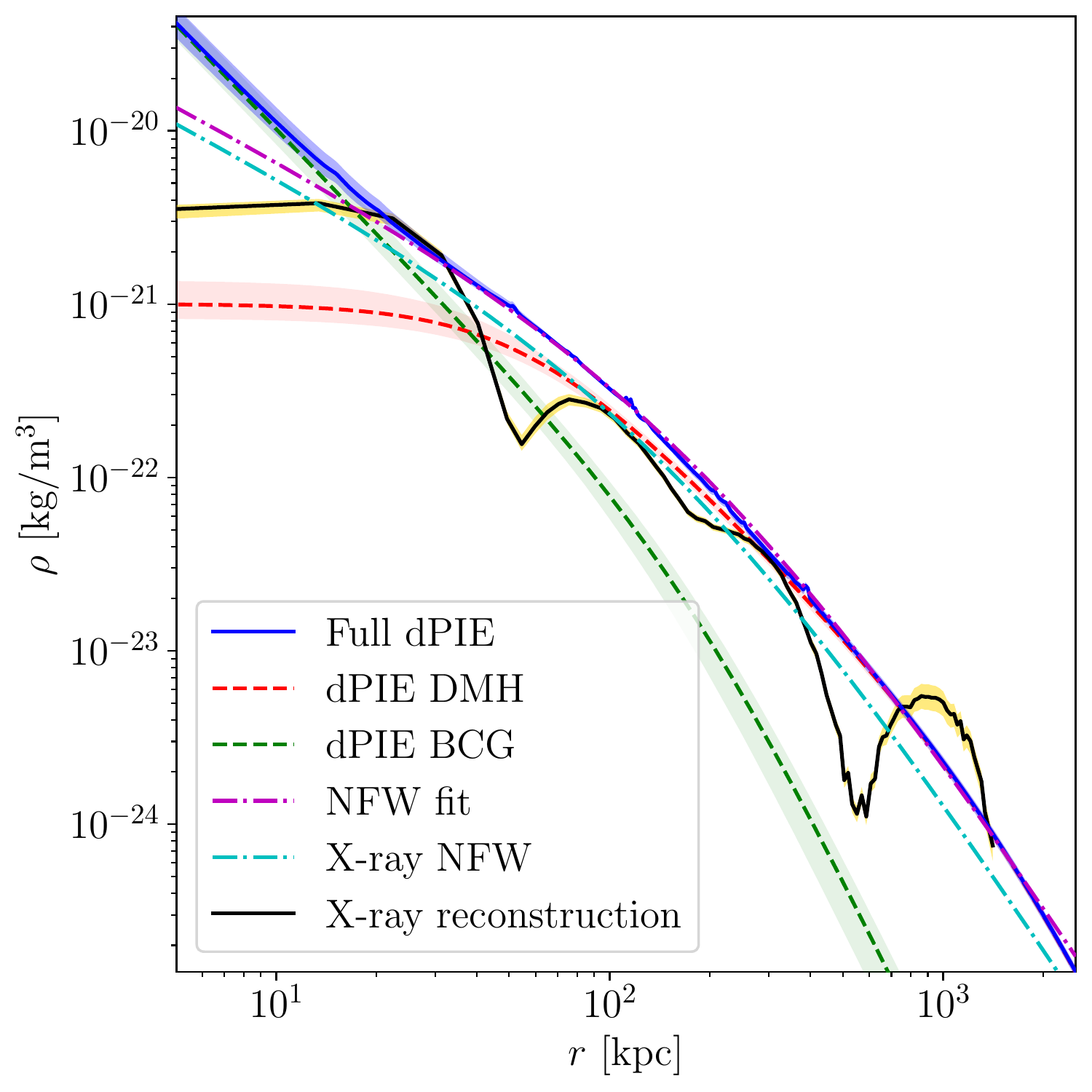}}
\end{minipage}
\begin{minipage}{0.48\textwidth}
\centerline{\includegraphics[width=1\textwidth]{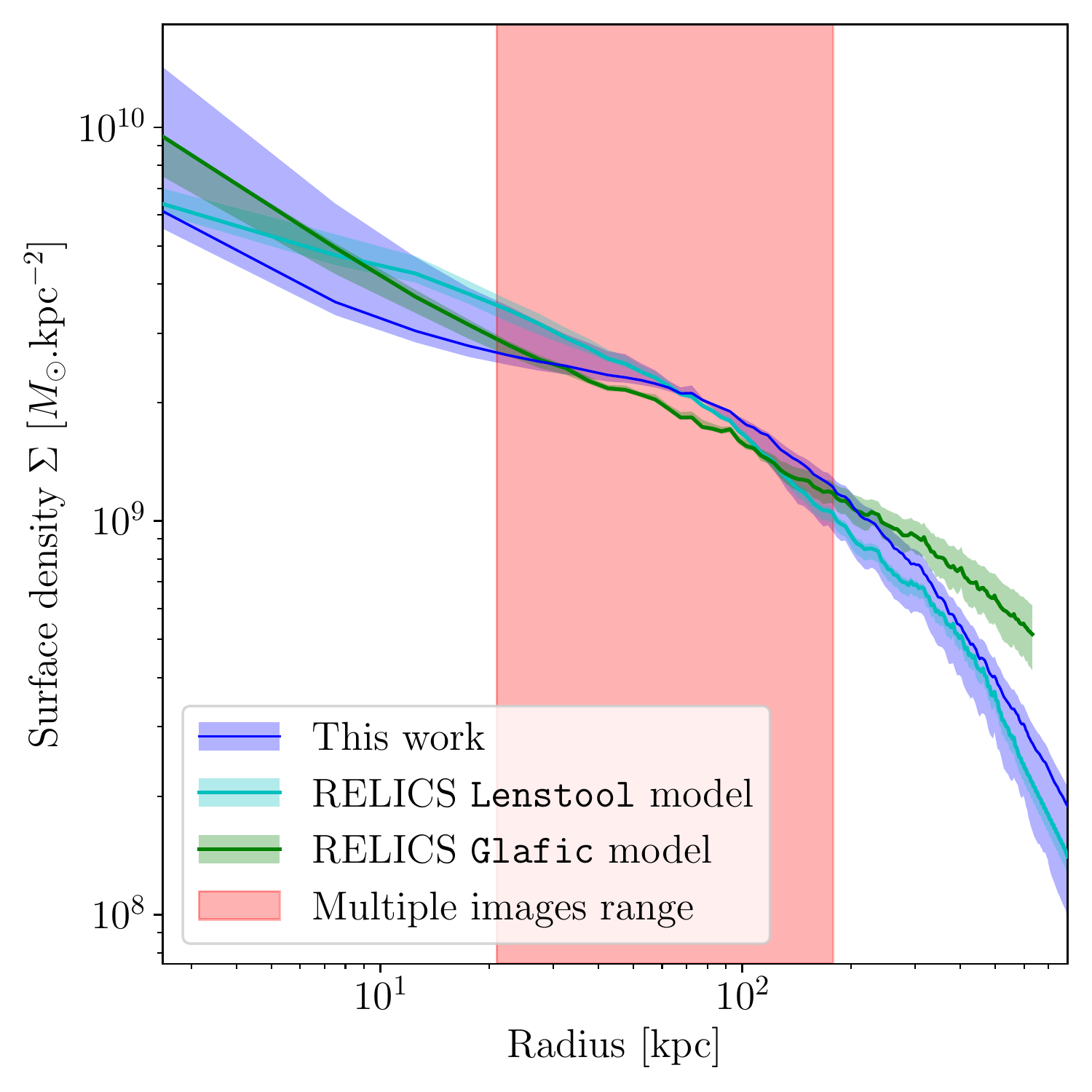}}
\end{minipage}
\hfill
\begin{minipage}{0.48\linewidth}
\centerline{\includegraphics[width=1\textwidth]{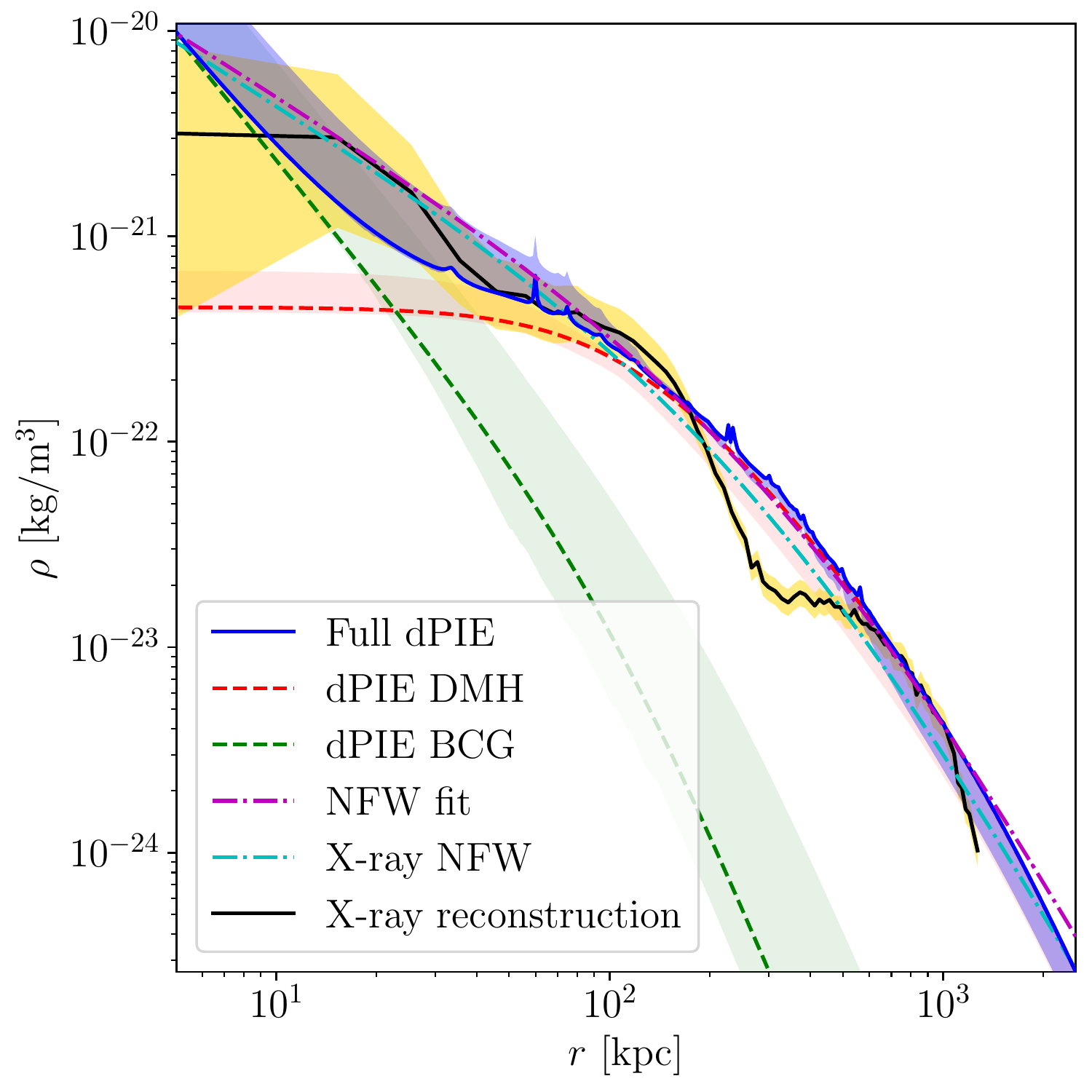}}
\end{minipage}
\caption{\textit{Top row}: Cluster MACS\,J0242. \textit{Left panel:} Surface mass density profile derived from the best-fit mass model.
Shaded regions show the $68 \%$ confidence interval. We display in red the range of the multiple images, and thus the regions in which the constraints are the most stringent. --
\textit{Right panel:} Volume mass density. The reconstruction of the \textit{XMM-Newton} observations are shown in black, given with $1\sigma$ error bars in yellow. The green and red curves -- with error bars -- represent respectively the BCG and DM halo reconstructions, and the full cluster is shown in blue. The magenta dashed line represents the NFW fit of the total density from \texttt{Lenstool} reconstruction -- all galaxies and DM halo. The cyan line shows the fit to the X-ray data.
\textit{Bottom row}: Cluster MACS\,J0949.
\textit{Blue:} Our model, with $68 \%$ confidence interval.
\textit{Cyan:} \texttt{Lenstool} model from RELICS. We note that error bars were obtained on a different sample (2,000 realisations for our model, 100 for RELICS). \textit{Green:} \texttt{Glafic} RELICS model, realised under the same conditions. \textit{Red:} region of the multiple images constraints. -- \textit{Right panel:} Volume mass density. The reconstruction of the \textit{XMM-Newton} data is shown in black, given with $1\sigma$ error bars in yellow. The green and red curves represent respectively the BCG and DM halo reconstruction, and the full cluster is shown in blue. The magenta dashed line represents the NFW fit to the \texttt{Lenstool} reconstruction. The cyan line shows the fit to the X-ray data.}
\label{fig:all_density_plots}
\end{figure*}

The geometry of the cluster is typical of a relaxed cool-core cluster.
The density profiles peak in the centre, and the transition between the BCG and the DM halo appears to be very smooth as illustrated in Fig.\,\ref{fig:all_density_plots}. No other significant structure are identified.
Figure\,\ref{fig:all_density_plots} shows the surface density profile, $\Sigma$, and includes a $68 \%$ confidence interval around the best contours, as a function of the distance to the cluster centre.
The inner part of the profile, $R \lesssim 50$\,kpc, is dominated by the BCG potential, while at larger radii, the dark matter halo takes over.
This pivot scale of about 50\,kpc corresponds to the core radius of the DM halo, and the separation between the two different regimes of the dPIE potential.
However, disentangling the potential influence of the BCG and the DM of the halo would require a much finer study of the stellar mass distribution of the BCG with a spectral energy distribution (SED) fit, which is beyond the scope of this article.

We find the total density profile (baryonic and dark matter) of MACS\,J0242 to be well fitted by a Navarro-Frenk-White profile \citep[NFW, see][]{1996ApJ...462..563N} in the region between 20 and 1000\,kpc.
We limit the reconstruction to radii $r \geq 20$\,kpc as the Kron-like magnitude radius of the BCG is about $10$\,kpc, and we attempt to limit the influence of stellar physics within the fit. 
In order to compare it to the NFW fit of cluster MACS\,J0949, we arbitrarily take $20$\,kpc to be a good compromise of strong lensing potential reconstruction without stellar physics contamination.
For regions $r > 200$\,kpc, the cluster-scale DM halo should dominate the whole matter distribution. 
As the DM halo dPIE parameters $\rho_0$ and $r_{\mathrm{core}}$ are well constrained through strong lensing, this region beyond multiple images constraints and below the cut-off radius $r_{\mathrm{cut}}$ is expected to be well represented by a NFW profile.
With NFW parameters $\rho_S = 3.42 \times 10^{-22}$\,kg.m$^{-3}$ and $r_S = 209.9$\,kpc, we find a reduced $\chi^2 = 1.11$.

In order to compare our results to the X-ray data, we extrapolate the masses $M_{\Delta, c}$ comprised within an overdensity $\Delta$ using 
\begin{equation}
    R_{\Delta} = \left\{ R \suchthat \frac{M(< R)}{\frac{4}{3} \pi R^3} = \Delta \cdot \rho_c (z) \right\},
    \label{eq:R_Delta}
\end{equation}
where $\rho_c$ is the critical density at the cluster redshift, and $M(< R)$ the total mass enclosed within a given radius, $R$.
At large radii ($R > 200$\,kpc), the strong lensing mass reconstruction only provides an estimate of the true mass distribution as there is no strong lensing constraints to precisely and accurately estimate the mass distribution in the outskirts. It therefore only provides a pure extrapolation of the inner core mass distribution, and only a weak-lensing analysis would provide a precise mass estimate in this region of the cluster, however this is beyond the scope of this analysis.
We also compute $M_{2D}(R<200$\,kpc), the integrated mass within a radius of 200\,kpc. This mass is a direct output of the lensing mass reconstruction.
These values are all listed in Table\,\ref{tab:mass_summarised}.

\subsubsection{MACS\,J0949 model}
\label{subsubsec:mass_model_m0949}

\begingroup
\renewcommand{\arraystretch}{1.2}
\begin{table}
	\centering
	\caption{List of the multiple images detected with VLT/MUSE in MACS\,J0949. We here list their ID, coordinates, R.A. and Decl. given in degrees (J2000), and their measured spectroscopic redshift $z$. Values within brackets were obtained after \texttt{Lenstool} redshift optimisation.}
	\label{tab:spectro_mul_m0949}
	\begin{tabular}{lccc}
	    \hline
		\hline
		Id. & R.A. & Dec. & $z$\\
		\hline
        1.1 & 147.4683753 & 17.11409360 & 4.8902\\
        1.2 & 147.4738000 & 17.11754490 & 4.8902\\
        1.3 & 147.4561230 & 17.11911410 & 4.8902\\
        1.4 & 147.4687438 & 17.12369520 & 4.8902\\
        1.5 & 147.4668972 & 17.12016960 & 4.8902\\
        2.1 & 147.4687829 & 17.11396160 & 4.8844\\
        2.2 & 147.4735428 & 17.11690610 & 4.8844\\
        2.3 & 147.4560463 & 17.11877380 & 4.8844\\
        2.4 & 147.4685346 & 17.12338060 & 4.8844\\
        3.1 & 147.4702800 & 17.11513600 & [$4.85_{-0.70}^{+1.52}$] \\
        3.2 & 147.4714400 & 17.11579400 & [$4.85_{-0.70}^{+1.52}$] \\
        4.1 & 147.4630587 & 17.10291430 & [$3.76_{-0.80}^{+1.57}$] \\
        4.2 & 147.4642781 & 17.10251570 & [$3.76_{-0.80}^{+1.57}$]\\
        4.3 & 147.4663104 & 17.10264970 & [$3.76_{-0.80}^{+1.57}$] \\
        5.1 & 147.4631754 & 17.10292500 & [$3.63_{-0.74}^{+1.67}$] \\
        5.2 & 147.4641921 & 17.10257190 & [$3.63_{-0.74}^{+1.67}$] \\
        5.3 & 147.4664329 & 17.10269780 & [$3.63_{-0.74}^{+1.67}$] \\
        6.1 & 147.4633639 & 17.10469208 & [$3.57_{-1.08}^{+0.35}$] \\
        6.2 & 147.4644174 & 17.10467818 & [$3.57_{-1.08}^{+0.35}$] \\
        6.3 & 147.4665100 & 17.10432399 & [$3.57_{-1.08}^{+0.35}$] \\
		\hline
		\hline
	\end{tabular}
\end{table}
\endgroup

In MACS\,J0949, we identified several objects located behind the cluster with the MUSE observations. However most of them appear to be singly lensed.
Through the techniques exposed in Sect.\,\ref{sec:Analysis}, we detected a multiple image system in the MUSE field at redshift $z = 4.8902$. 
This system 1 is composed of five multiple images, including four in the field, and one counterpart 1.3 located outside the MUSE field of view, and detected in the \textit{HST} imaging.
We also detect a fifth image, image 1.5, located close the BCG of the cluster.
Images 1.4 and 1.5 (see Fig. \ref{fig:astro_im_m0949}), straddling the central critical curve of the cluster, allow to set stringent constraints on the inner slope of the mass density profile \citep[as exhibited in][]{1992grle.book.....S,2013ApJ...765...25N, 2017A&A...607A..93C}.

Careful consideration of the \textit{HST} images allowed us to detect secondary, fainter emission knots for four multiple images in system 1 -- all except the central one which is hidden by the emission of the BCG. This is shown in Fig.\,\ref{fig:m0949_spectro_multimages}. The MUSE spectroscopic analysis of these three images which compose system 2 shows a faint Ly-$\alpha$ peak for all of them, allowing us to measure a redshift of $4.8844$, very close to that of system 1.
We interpret system 2 either as part of the same galaxy, or a companion galaxy of system 1's source. The Ly-$\alpha$ halo of system 1 extends, and the potential secondary peak emission coincides with system 2 emission knots.
We include 4 multiple images of system 2 as additional constraints to our mass model, the fifth image being demagnified we restrain ourselves from including it in our mass model.
The coordinates and redshifts of the multiply imaged systems are given in Table\,\ref{tab:spectro_mul_m0949}. We give a list of the singly imaged objects in Appendix\,\ref{sec:spectroscopic_detections_of_interest}.

\begin{figure}
    \centering
    \includegraphics[width=\columnwidth]{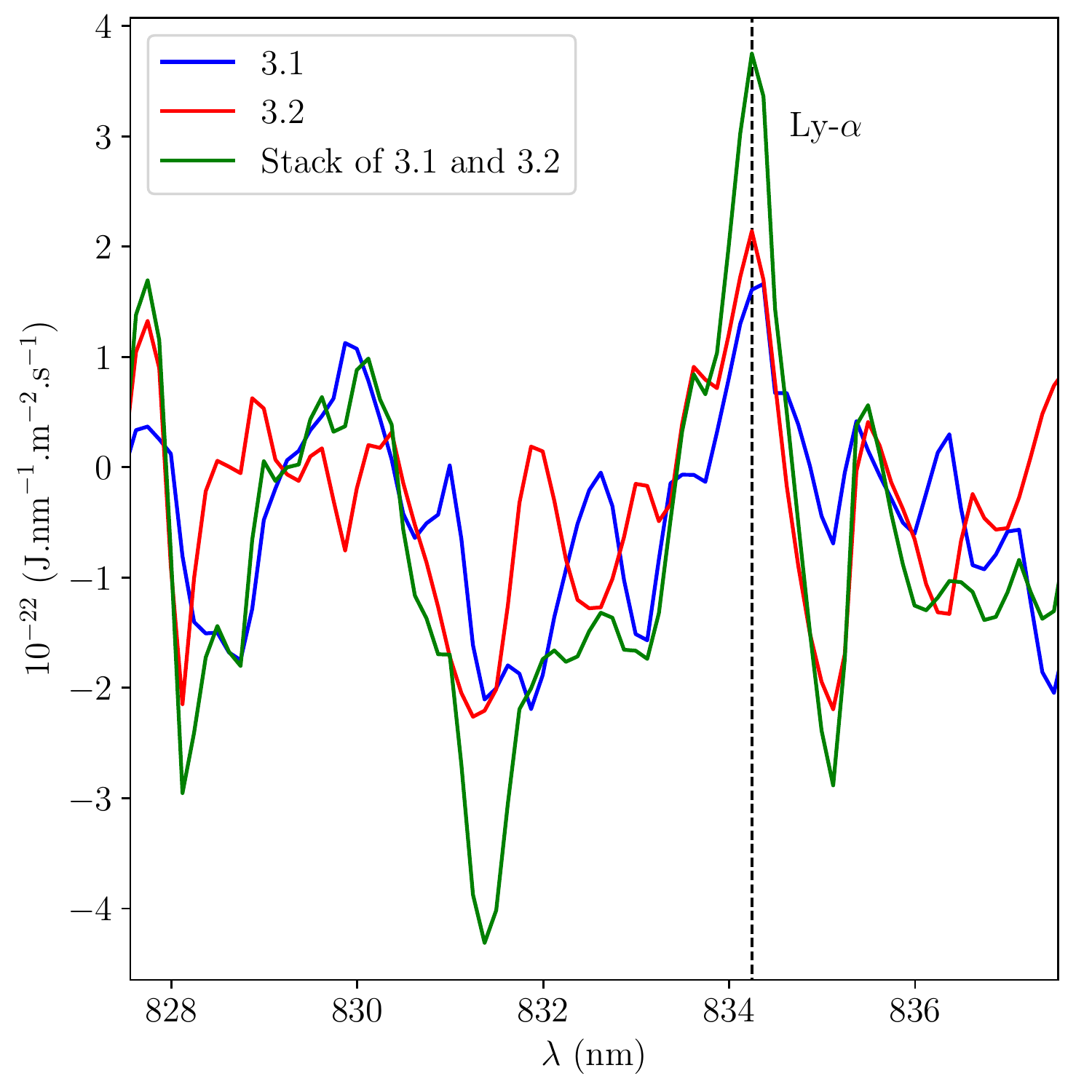}
    \caption{Spectra of images 3.1 and 3.2 of cluster MACS\,J0949 obtained by VLT/MUSE. We can observe a faint signal, possibly Ly-$\alpha$. \textit{Blue}: spectrum of 3.1; \textit{Red}: spectrum of 3.2; \textit{Green}: summed spectra. The redshift measured would be of $5.8658$. However, the confidence level of our measurements is low due to high sky noise at this wavelength.}
    \label{fig:Ly-alpha_sys3_m0949}
\end{figure}

The inspection of \textit{HST} images also led to the discovery of system 3, composed of two multiple images. These faint detections in the South of the cluster were equally present in the MUSE field. A faint and \textit{a priori} inconclusive detection of Ly-$\alpha$ -- see Fig.\,\ref{fig:Ly-alpha_sys3_m0949} -- is consistent with the redshift optimisation of this system using only system 1, or 1 and 2 as constraints.
We therefore conclude that this system's redshift is 5.8658. However the stack of the spectra presents a S/N ratio $< 2$, and the MUSE data are sensible to sky perturbations in the speculated Ly-$\alpha$ bandwidth. We therefore decide not to use this as a redshift constraint, but to let the redshift free during the model optimisation.

At last, we detect three candidate multiply lensed images in the South of the \textit{HST} field of view, in a region not covered by the MUSE observations. We included these three candidate systems 4, 5 and 6 in our mass model, letting their redshifts as free parameters.
Their detection supposes the presence of a Southern halo as described in Sect.\,\ref{sec:Mass-modelling}.
For systems 3, 4, 5 and 6, our best fit mass model gives the respective redshifts: $4.85_{-0.70}^{+1.52}$, $3.76_{-0.80}^{+1.57}$, $3.63_{-0.74}^{+1.67}$ and $3.57_{-1.08}^{+0.35}$.

Similarly to MACS\,J0242, we model the mass distribution of the cluster scale halo and the BCG galaxy separately.
The best-fit mass model parameters are listed in Table\, \ref{tab:best_model}, and gives a $rms$ of 0.15\arcsec.
The addition of an external shear component does not improve the mass model, and gives a $rms$ of $0.16\arcsec$.
In a similar fashion to MACS\,J0242, although the degeneracy between the cluster scale halo and the BCG is still present, the BCG optimisation converges.
The $rms$ is particularly small which may be explained by the lack of constraints in our model.
Indeed, as shown in e.g. \citet{Johnson_2016}, a larger number of constraints may increase the value of the $rms$ but could also improve the accuracy of the model.
Similarly to MACS\,J0242, we compute integrated and 3D masses for MACS\,J0949. These are listed in Table\,\ref{tab:mass_summarised} and discussed further in Sect.\,\ref{sec:Discussion}.

We compare our model of MACS\,J0949 to the two publicly available models from the RELICS collaboration\footnote{\url{https://archive.stsci.edu/prepds/relics/}}. Comparing the surface density profiles, we find a $1\sigma$ agreement between the model presented in this article and the \texttt{Lenstool} RELICS model as can be seen in Fig.\,\ref{fig:all_density_plots}.
As for the RELICS model obtained using the \texttt{Glafic} lensing algorithm \citep[presented in][]{2010ascl.soft10012O}, its density profile is in agreement with our model, although the most stringent constraints (in the $R \in [40, 100]$\,kpc region) yield a slightly smaller surface density.
The overall profile from the \texttt{Lenstool} RELICS public release model presents a flatter density profile and an excess in mass after 80\,kpc (coincidental with the Einstein radius of system 1). This could be partially explained by the more massive structure in the South of the cluster, which is slightly offset from the South bright galaxy surrounded by systems 4, 5, and 6 as mentioned before ($M_{2D}(< 100$\,kpc) $= 13.02 \times 10^{12} \Msol$ compared to $M_{2D}(< 100$\,kpc) $= 7.65 \times 10^{12} \Msol$ for our model). 
We report a very good agreement between the measured spectroscopic redshift obtained from MUSE observations with the photo-$z$ used by the RELICS team (obtained through private communication with K. Sharon).
Our model presents a significantly lower $rms$ of $0.15\arcsec$, in comparison to $0.58\arcsec$.

The reconstructed mass distribution appears to be more elliptical than the X-ray surface brightness obtained with \textit{XMM-Newton} as shown in Fig.\,\ref{fig:astro_im_m0949}. The 3D density profile is presented in Fig.\,\ref{fig:all_density_plots}. It confirms the inflexion point in the density profile at $r \simeq 100$\,kpc, and therefore suggests that the cluster is still undergoing a relaxing phase.
The NFW profile fit in the $r \in [20, 1000]$\,kpc region yields NFW parameters $\rho_S = 1.23 \times 10^{-22}$\,kg.m$^{-3}$, $r_S = 405.5$\,kpc, for a reduced $\chi^2 = 1.90$. The quality of this fit is thus not comparable to that of cluster MACS\,J0242, mostly due to the flatter density profile in the $R \in [40, 100]$\,kpc region.

Looking at the galaxy distribution within the cluster,
we observe four bright and massive galaxies, of comparable magnitude to the BCG\footnote{The maximum magnitude separation between these five galaxies being 0.29 on the reference band ACS/F814W.}.
We could extrapolate all of these bright galaxies to have been the BCG of former galaxy clusters, which would have merged with MACS\,J0949 in the past.
However, the X-ray observations show a diffuse emission centred on the BCG and thus do not provide any evidence of recent merger events.
Therefore, our analysis strongly suggests a unique dominant cluster scale dark matter component.
Nonetheless, we stress that the magnitude gap between the BCG and the second-brightest cluster galaxy in MACS\,J0242 is much larger than in MACS\,J0949. According to \citet{2017MNRAS.471.2022T}, this is an additional argument to claim that the former cluster is more relaxed, and that MACS\,J0949 went through a recent merging event.

Our interpretation of the dynamical state of MACS\,J0949 and its lensing power could be further constrained with additional spectroscopic or imaging observations. The clear identification of the spectroscopic redshift of system 3, and of additional systems would particularly assist constraining the dark matter halo ellipticity, core radius and velocity dispersion.

\subsubsection{Relensing in MACS J0949}

\begin{figure}
    \centering
    \includegraphics[width=\columnwidth]{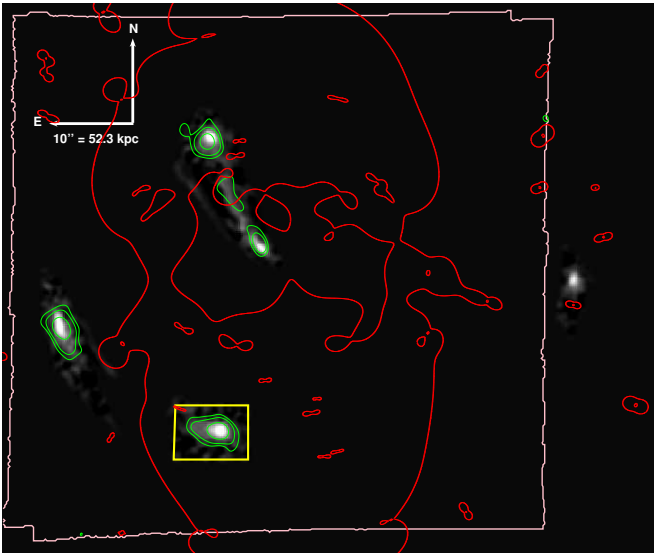}
    \caption{
    MACS\,J0949 reconstruction of the full image plane of system 1 from the unique extended emission images 1.1 and 2.1. Their region, highlighted with the yellow box is cut out and deprojected into the source plane, and casted back in the image plan to produce the full system. We clearly observe a continuous emission between the North-East image 1.4 and the central one 1.5.
    We display in green the contours of the Ly-$\alpha$ extended emission from the VLT/MUSE narrow-band image centred at 715.869\,nm and 1.625\,nm wide, showing the four detected multiple images of system\,1, and three of system 2 (see Fig.\,\ref{fig:astro_im_m0949} for more details).
    The last images 1.3 and 2.3 of these systems are located outside of the VLT/MUSE field of view. The critical lines are displayed in red, for redshift $z = 4.8902$ of system 1. The pink overlay represents the MUSE narrow-band contours.}
    \label{fig:MUSE_detection_m0949}
\end{figure}

On Fig.\,\ref{fig:MUSE_detection_m0949}, we display the extracted emission of images 1.1 and 2.1 detected in MACS\,J0949 from the MUSE narrow-band centred on $\lambda = 715.869$\,nm within a yellow box.
In order to verify the robustness of the lensing model of MACS\,J0949, we
then infer the emission in the source plane ($z=4.8902$), before projecting it back to the image plane with our lens model, to obtain a re-lensed prediction.

The other multiple images on the MUSE field, 1.2, 1.4, 1.5, 2.2 and 2.4 are correctly predicted. Their Lyman-$\alpha$ detections are also listed in Table\,\ref{tab:spectro_mul_m0949}.
Images\,1.4 and 1.5 emission appear to be connected.
This is simply due to the extended source emission of system 1 and 2, as a number of faint multiple images of system 2 are predicted between 1.4 and 1.5, in agreement to the MUSE observations on the narrow-band.

\subsection{Stellar mass estimate}

The strong lensing analyses are giving us an estimate of the total mass enclosed in each clusters.

We  further compare our strong lensing mass with an estimate stellar mass.
We use the reference cluster members catalogue magnitudes described in Sect.\,\ref{sec:Analysis}, converted into K-band luminosity $L_K$\footnote{We take the K-band reference here to be the \href{http://svo2.cab.inta-csic.es/theory/fps/index.php?id=KPNO/Flamingos.Ks&&mode=browse&gname=KPNO&gname2=Flamingos\#filter}{KPNO Flamingos Ks} filter.}, and use it as a proxy for stellar mass. 
For the scaling relations we refer the reader to \citet{2002astro.ph.10394H, 2006ApJ...650L..99L}.
These catalogues were established over the entire observable clusters, although the faintest galaxies were cut out beyond distances of 40\arcsec~from the centre.

Once the $L_K$ catalogue established, we adapt the Salpeter initial mass function, and use the mass-to-light relationship for red quiescent galaxies derived by \cite{2007A&A...476..137A} on the SWIRE-VVDS-CFHTLS surveys, based on the \citet{2003MNRAS.344.1000B} stellar population models:
\begin{equation}
    \log_{10}\left[ \frac{M_{\star}}{\Msol} \frac{L_{\mysun}}{L_K} \right] = a z + b,
    \label{eq:Arnouts2007}
\end{equation}
given the parameters $\{a, b\} = \{-0.18 \pm 0.04, +0.07 \pm 0.04\}$.
While we acknowledge our studied clusters are within a redshift range presenting large uncertainties in the relationship presented in \citet[][see Fig.\,9]{2007A&A...476..137A}, we refer the reader to the detailed comparison made in Appendix D, Fig.\,28 of \citet{2010ApJ...709..644I}.
Although the former appears to overestimate the stellar mass by an average 0.2 dex for red sequence galaxies, it also appears to be reasonably well calibrated for $z \in [0.3, 0.4]$.
We present the inferred stellar masses for both clusters in Table\,\ref{tab:mass_summarised}. 

\begingroup
\renewcommand{\arraystretch}{1.5}
\begin{table*}
	\centering
	\caption{Mass and radius measurements for MACS\,J0242 and MACS\,J0949. All error bars show a $68 \%$ confidence interval.
	We here list $M_{\star}$, the stellar mass, $M_{\mathrm{2D}} (R < 200\,\mathrm{kpc})$, the mass distribution obtained in projection on the plane of the cluster, within a radius of 200\,kpc, and $M_{\Delta}$ and $R_{\Delta}$, defined in eq.\,(\ref{eq:R_Delta}). Masses are given in $10^{14} \Msol$ and distances in kpc.
	The X-ray masses are following the NFW fit.}
	\label{tab:mass_summarised}
\makebox[\textwidth][c]{
    \begin{tabular}{lcccc}
\hline
\hline
   & \multicolumn{2}{c}{MACS J0242} & \multicolumn{2}{c}{MACS J0949} \\
\cmidrule(lr){2-3}\cmidrule(lr){4-5}
Mass ($10^{14} \Msol$) & Lensing & X-ray & Lensing & X-ray\\ \midrule
   $M_{\star}$  & \multicolumn{2}{c}{$0.065 \pm 0.006$} & \multicolumn{2}{c}{$0.139 \pm 0.014$} \\
	$M_{\mathrm{2D}} (R < 200 \mathrm{kpc})$  & $1.667_{-0.052}^{+0.032}$ & $1.163_{-0.039}^{+0.036}$ & $1.996_{-0.199}^{+0.051}$ & $1.635_{-0.072}^{+0.065}$ \\
	$M_{2500}$ & $3.113_{-0.200}^{+0.160}$ & $1.875^{+0.070}_{-0.069}$ & $5.621_{-0.942}^{+0.122}$ & $3.439^{+0.281}_{-0.266}$\\
	$M_{1000}$ & $4.628_{-0.342}^{+0.289}$ & $2.695^{+0.122}_{-0.121}$ & $8.848_{-2.215}^{+0.000}$ & $5.547^{+0.778}_{-0.693}$\\
	$M_{500}$  & $5.954_{-0.455}^{+0.400}$ & $3.379^{+0.168}_{-0.168}$ & $11.483_{-3.417}^{+0.000}$ & $7.429^{+1.310}_{-1.137}$\\
	$M_{200}$  & $7.748_{-0.598}^{+0.538}$ & $4.343^{+0.238}_{-0.237}$ & $14.790_{-4.824}^{+0.000}$ & $10.165^{+2.234}_{-1.799}$\\
\hline
\hline
   & \multicolumn{2}{c}{MACS J0242} & \multicolumn{2}{c}{MACS J0949} \\
\cmidrule(lr){2-3}\cmidrule(lr){4-5}
	Radius (kpc) & Lensing & X-ray & Lensing & X-ray\\
	\hline
	$R_{2500}$ & $541.2_{-11.9}^{+9.1}$ & $466^{+5}_{-6}$ & $641.7_{-38.1}^{+4.6}$ & $555^{+15}_{-15}$ \\
	$R_{1000}$ & $838.3_{-21.2}^{+17.1}$ & $713^{+11}_{-11}$ & $1013.2_{-92.8}^{+0.0}$ & $884^{+39}_{-39}$ \\
	$R_{500}$  & $1148.7_{-30.1}^{+25.2}$ & $969^{+16}_{-16}$ & $1392.4_{-154.6}^{+0.0}$ & $1227^{+68}_{-66}$ \\
	$R_{200}$  & $1702.0_{-45.0}^{+38.5}$ & $1430^{+26}_{-27}$ & $2056.1_{-253.5}^{+0.0}$ & $1849^{+126}_{-116}$ \\
	\hline
	\hline
\end{tabular}
}
\end{table*}
\endgroup

In order to have a theoretical reference, we compare our estimates with the stellar mass predicted using the formula derived by \citet{Giodini_2009}. 
This relationship, established for poor clusters, with redshifts $0.1 \leq z \leq 1$, relates the total mass of the cluster to its stellar fraction ($M_{\star}/M_{500}$ here) using the relation:
\begin{equation}
   f^{\star}_{500} = 0.05^{+0.001}_{-0.001} \left( \frac{M_{500}}{5 \times 10^{13} \Msol} \right)^{-0.37 \pm 0.04}.
    \label{eq:stellar_fraction_Giodini}
\end{equation}
Let us notice the high ($\sim 50\%$) logarithmic scatter in the data fitting this relationship. 
As this relationship was established using X-ray measurements of $M_{500}$, and that strong lensing is not a direct probe of this value, we use the NFW reconstruction obtained through X-ray for the $M_{500}$ values (see Fig.\,\ref{fig:all_density_plots}).

For MACS\,J0242, the field of view considered is quite large (DES: $182\arcsec$), as we consider all galaxy in \textit{HST}/WFPC2 or DES, and thus our cluster member catalogue is assumed to be relatively complete.
We measure a stellar mass $M_{\star} = (6.484 \pm 0.615) \times 10^{12} \Msol$ for MACS\,J0242. Let us notice these error bars are only associated to the error on the measured magnitude and the parameters $a$ and $b$ eq.\,(\ref{eq:Arnouts2007}).
We obtain a difference between our measured value and the predicted value of $M_{\star, {\rm Giodini}} = (8.332 \pm 1.128) \times 10^{12} \Msol$.
We may explain this discrepancy by the variable conditions for selecting a galaxy within the galaxy catalogue. Indeed, the field of view being different between WFPC2, ACS and DES, as well as the poorer imaging quality of the latter instrument, we expect our error bars to be far larger than those computed given the error on the measured magnitude.

For MACS\,J0949, we require that a galaxy is detected in either \textit{HST}/ACS or \textit{HST}/WFC3 to include it in the final catalogue.
Because the field of view of WFC3 is smaller than that of ACS, a large number of selected cluster member galaxies are weakly constrained, as ACS only contains two bands here.
This method is adapted to our lensing analysis, the main goal of this paper, as galaxies far from the cluster centre are particularly important to constrain the southern halo.
However, when considering the stellar content of the cluster, we might be selecting too many galaxies.
Our analysis yields $M_{\star} = (1.392 \pm 0.137) \times 10^{13} \Msol$.
Similarly to MACS\,J0242, we compare our measurement with the predicted value following the \citet{Giodini_2009} formula. We obtain a stellar mass $M_{\star, {\rm Giodini}}=(1.369 \pm 0.302) \times 10^{13} \Msol$.
This difference, however small, can give us an estimate of the overestimation of our cluster member catalogue.
We summarise the estimated stellar fractions for both clusters, $f^{\star}_{500} = M_{\star}/M_{500}$, as well as the predicted values with the \cite{Giodini_2009} formula in Table\,\ref{tab:stellar_fraction_compared}.

\begin{table}
	\centering
	\caption{Comparison between the star fractions $f^{\star}_{500} = M_{\star}/M_{500}$ measured with this work, and the predictions from the \citet{Giodini_2009} formula. $M_{500}$ is taken to be the NFW X-ray extrapolated value. All results are in percentage.
	}
	\label{tab:stellar_fraction_compared}
	\begin{tabular}{lcc}
	    \hline
		\hline
		$f^{\star}_{500} (\%)$ &  MACS\,J0242 & MACS\,J0949\\
		\hline
		This work  & $1.919 \pm 0.205$ & $1.873 \pm 0.360$\\
		Prediction & $2.466 \pm 0.334$ & $1.842 \pm 0.407$\\
		\hline
		\hline
	\end{tabular}
\end{table}

\subsection{X-ray analysis}
\label{subsec:X-ray_analysis}

\subsubsection{Analysis procedure}

We used the X-COP analysis pipeline \citep{ghirardini19} to analyse the data and compute the hydrostatic mass profiles of the two systems. We extracted X-ray photon images in the [0.7-1.2] keV band, which maximises the signal-to-background ratio. To estimate the non X-ray background, we used the unexposed corners of the MOS detectors to estimate the cosmic-ray-induced flux at the time of the observations. The difference between the scaled high-energy count rates inside and outside the field of view were then used to estimate the residual soft proton contribution, which was next modelled following the method described in \citet{ghirardini18}. To determine the spectroscopic temperature profile of the two systems, we extracted spectra in logarithmically spaced concentric annuli centred on the surface brightness peak. The sky background emission was measured in regions located well outside of the cluster's virial radius and described by a three-component model including the cosmic X-ray background, the local hot bubble, and the galactic halo. The sky background spectrum was then rescaled appropriately to the source regions and added as an additional model component. Finally, the source spectrum was modelled by a single-temperature APEC model \citep{apec} absorbed by the Galactic $N_{H}$, which was fixed to the HI4PI value \citep{HI4PI}. 

\subsubsection{Hydrostatic mass reconstruction}

We used the publicly available Python package \texttt{hydromass}\footnote{\href{https://github.com/domeckert/hydromass}{https://github.com/domeckert/hydromass}} \citep[][]{2022A&A...662A.123E} to deproject the X-ray data and recover the mass under the hypothesis of hydrostatic equilibrium. The X-ray surface brightness and spectroscopic temperature profiles are fitted jointly using a NFW profile to recover the X-ray mass profile. The technique employed here is similar to the method described in \citet{ettori19}, in which the gas density profile and the parametric mass profile are used to integrate the hydrostatic equilibrium equation and predict the 3D pressure and temperature profiles. The 3D temperature profile is then projected along the line of sight using spectroscopic-like weights \citep{mazzotta04} and adjusted onto the observed spectroscopic temperature profile. The model temperature and gas density profiles are convolved with the \textit{XMM-Newton} PSF to correct for the smearing introduced by the telescope's spatial resolution, in particular in the cluster's central regions.

\subsubsection{MACS\,J0242}
MACS\,J0242 exhibits all the features of a relaxed, cool-core cluster. Its X-ray morphology is regular and it shows a pronounced surface brightness peak, a central temperature drop, and a metal abundance peak in its core.
The dynamical state of the cluster is best gauged from the X-ray emission, but the optical emission lines of the BCG is an additional, relatively faithful tracer of the presence of a cool core.
The NFW mass reconstruction returns a mass $M_{500} = (3.4 \pm 0.2)\times 10^{14}\,M_\odot$. 
In order to compare it directly to the lensing mass where multiply imaged systems yield important constraints, we project the NFW density in 2D, and compute $M_{\mathrm{2D}} (<200$\,kpc) $= 1.163_{-0.039}^{+0.036} \times 10^{14} \Msol$.
For an average temperature of 4.5\,keV, this is in agreement with the expectations of mass-temperature relations \citep[e.g.][]{lovisari20}. The cluster appears to be highly concentrated, with a fitted NFW concentration $c_{200} = 8.2 \pm 0.5$.
At 200\,kpc, X-ray observations suggest the gas fraction to be $f_{g, \mathrm{200\,kpc}} = 0.115^{+0.003}_{-0.004}$.
The ellipticity of the cluster obtained with our lensing mass model is not recovered by the X-ray analysis, as it presents a spherical surface brightness.
The ICM has its own dynamics and thus is not expected to present a similar ellipticity to the total density of matter.
The discrepancy between the ICM and DM halo ellipticity is documented in e.g. \citep{2003ApJ...585..151L, 2008ApJ...681.1076D, 2012ApJ...755..116L, 2018ApJ...860..104U, 2022A&A...663A..17S}. It stems from the collisional character of baryons, allowing the ICM to geometrically relax faster than the cold dark matter halo counterpart, non-collisional.

\subsubsection{MACS\,J0949}
MACS\,J0949 exhibits a regular X-ray morphology with no obvious large substructure. However, its brightness distribution is relatively flat, it shows a high central entropy and central cooling time, and no temperature drop in its core. Therefore, MACS\,J0949 is not a relaxed cool-core cluster, but its regular morphology indicates that it is not strongly disturbed either. Such properties are typical of post-merger clusters in the process of relaxation after a merging event. The hydrostatic mass profile is well described by an NFW model with $c_{200} = 5.3_{-1.0}^{+1.3}$ and $M_{500} = 7.4_{-1.2}^{+1.4}\times 10^{14} M_\odot$.
The NFW projected mass yields $M_{\mathrm{2D}} (<200$\.kpc) $= 1.635_{-0.072}^{+0.065} \times 10^{14} \Msol$.
Its hydrostatic gas fraction $f_{g,500} = 0.155_{-0.014}^{+0.016}$ is consistent with the Universal baryon fraction \citep{Planck_XIII_2016}.
At 200\,kpc, the same gas fraction is measured at $f_{g, \mathrm{200\,kpc}} = 0.053^{+0.007}_{-0.006}$.
Similarly to MACS\,J0242, the X-ray signal does not present any ellipticity.

\section{Discussion \& Conclusion}
\label{sec:Discussion}

In order to reconstruct the mass distribution of strong lensing galaxy clusters MACS\,J0242 and MACS\,J0949, we have used the combination of imaging (\textit{HST}, DES) and spectroscopic (VLT/MUSE) surveys to detect respectively 6 and 2 spectroscopically confirmed multiple image systems. Adding to that, in MACS\,J0949, we identified four multiply imaged systems, without a confirmed spectroscopic redshift -- the spectroscopic emission line not fitting spectral templates convincingly enough, or the images being out of the VLT/MUSE field of view.
The imaging data, calibrated with the spectroscopic detections of cluster members, allowed to establish conservative cluster galaxy catalogues, of respectively 58 and 170 galaxies for MACS\,J0242 and MACS\,J0949.
We then established the strong lensing mass models of both galaxy clusters.
We modelled each individual galaxy with a dPIE profile, and included for each cluster a dPIE cluster-scale halo.
We present our main results as follows:
\begin{enumerate}
    \item The $rms$ on the multiple image positions for the best-fit models are respectively of $0.39\arcsec$ and $0.15\arcsec$, which is considered as a good quality indicator of the reconstruction. We found that adding a shear-field does not improve the quality of the reconstruction. We note that degeneracies between the BCG and the dark matter halo could hinder the lens model optimisations, and could thus affect our conclusion regarding the morphology of the dark matter distribution in these clusters \citep[see e.g.][]{2016A&A...588A..99L}.
    
    \item Using \textit{XMM-Newton} X-ray observations from \citet{2021A&A...650A.104C}, processed with the X-COP pipeline \citet{ghirardini19}, we compare the ICM to the reconstructed dark matter density. The combination of the lensing mass reconstructions with the X-ray analyses of the ICM and the VLT/MUSE spectroscopy shows that MACS\,J0242 is in a cool-core, relaxed dynamical state, compatible with a NFW profile, while MACS\,J0949 has a flat distribution between radii of 50 to 100\,kpc because it is still undergoing the relaxing process, being in a post-merger dynamical state.
    In particular, the hot gas fractions at $200$\,kpc of MACS\,J0242 and MACS\,J0949 are $f_{g, 200\,\mathrm{kpc}} = 0.115_{-0.004}^{+0.003}$ and $0.053_{-0.006}^{+0.007}$ respectively. We can for instance compare these results to those of \citet[][]{2018ApJ...864...98B}. In Fig.\,6, the authors present the cumulative hot gas fraction of each of the three clusters analysed. MACS\,J0416 is presented as a merging cluster, while MACS\,J1206 and Abell\,S1063 (RXC\,J2248) show a cool-core. These clusters have $f_{g, 200\,\mathrm{kpc}} \simeq 0.09$, $0.11$ and $0.13$ respectively, thus exhibiting the trend of more relaxed clusters displaying higher hot gas fraction values at 200\,kpc. This is an additional indication of the relaxed dynamical state of MACS\,J0242, and the post-merger state of MACS\,J0949.
    
    \item Converting the cluster member catalogue magnitudes into K-band luminosities, we used the \cite{2007A&A...476..137A} mass-to-light ratio relationship to extrapolate the stellar mass detected in both clusters. SED fitting should be performed to obtain a more precise measurement, but this is beyond the scope of this paper. We compare the obtained stellar masses of $M_{\star} = (6.48 \pm 0.62) \times 10^{12} \Msol$ and $(1.39 \pm 0.14) \times 10^{13} \Msol$ for MACS\,J0242 and MACS\,J0949 respectively to the predictions of \citet{Giodini_2009}, yielding respectively $(8.33 \pm 1.13) \times 10^{12} \Msol$ and $(1.37 \pm 0.30) \times 10^{13} \Msol$. Although not identical in the case of MACS\,J0242, this means our stellar mass estimates appear to be reasonable.
    
    \item We fit the \textit{XMM-Newton} observations to a NFW profile. Projecting this reconstruction, we can measure $M_{\mathrm{2D}} (<200$\,kpc), allowing for a direct comparison with the strong lensing model mass estimates. For MACS\,J0242, we measure $M_{\mathrm{2D}} (<200$\,kpc) $= (1.16 \pm 0.04) \times 10^{14} \Msol$ from the X-rays, to be compared to $1.67^{+0.03}_{-0.05} \times 10^{14} \Msol$ obtained from our strong lensing analysis. We obtain a sizeable $12.75\sigma$ difference between these two values.
    Discrepancies between the X-ray hydrostatic and lensing masses are common, and may be explained by the hydrostatic hypothesis bias, or by the presence of asymmetric structures along the line-of-sight. In the former case, the gas is not perfectly relaxed, and the thermal pressure only accounts for a fraction of the gravitational pressure. Thus, the hydrostatic mass would underestimate the true mass. Moreover, if there is a distribution of substructures or an elongation of the dark matter component along the line-of-sight,  the  projected lensing mass may overestimate the 3D mass. For instance, \citet{2015ApJ...806..207U} display a combination of both these scenarios.
    
    \item As for MACS\,J0949, we measure $M_{\mathrm{2D}} (<200$\,kpc) $ = (1.64 \pm 0.07) \times 10^{14} \Msol$ with the X-rays, to be compared with $2.00^{+0.05}_{-0.20} \times 10^{14} \Msol$ obtained with the strong lensing analysis. These values differ by $3.85\sigma$. The \texttt{Lenstool} and \texttt{Glafic} RELICS strong lensing models provide $M(R < 200\,\mathrm{kpc}) = 1.84_{-0.03}^{+0.03} \times 10^{14} \Msol$ and $M(R < 200\,\mathrm{kpc}) = 1.85_{-0.07}^{+0.08} \times 10^{14} \Msol$ respectively, in good agreement with our model. At last, we compare this latter value to the one obtained with the \textit{Planck} SZ data of $M_{2D}(< 200\,\mathrm{kpc}) = 1.59_{-0.00}^{+0.38} \times 10^{14} \Msol$ \citep[see][]{2022ApJ...928...87F}, assuming a NFW profile. This $1.49\sigma$ difference with the strong lensing value outlines a good agreement with our model.
\end{enumerate}

In order to compare cylindrical masses, we define $R_{10\%} = 0.1 R_{200,c}$.
For MACS\,J0242, with $R_{10\%} = 170.2_{-0.45}^{+0.39}$\,kpc, we obtain $M_{\mathrm{2D}} (<R_{10\%}) = (1.41 \pm 0.03) \times 10^{14} \Msol$ with our strong lensing analysis (for which $M_{200}$ is extrapolated). With $R_{10\%} = 143.0_{-2.6}^{+2.7}$\,kpc, we get $M_{\mathrm{2D}} (<R_{10\%}) = (8.06 \pm 0.21) \times 10^{13} \Msol$ with the X-rays NFW inferred profile, yielding ratios of $M_{\mathrm{2D}} (<R_{10\%}) / M_{200,c} = 0.181 \pm 0.014$ and $0.186 \pm 0.012$ respectively.
This allows us to characterise the ratios of masses measured in the centre and in the outskirts as quite close for X-ray and lensing, in spite of the remarkable difference between the mass measurements.
As the strong lensing inferred $M_{200}$ mass obtained here is an extrapolation at larger radii of a profile based on gravitational lensing occurring at $R < 200$\,kpc, we cannot claim the strong lensing ratios to be firmly established.
Nonetheless, the extrapolated lensing distribution appears to follow a profile similar to that of the X-rays, at different masses. 
We can compare this result to the ratios found by \citet{2018ApJ...864...98B} for three clusters exhibiting varied dynamical states (Abell\,S1063, MACS\,J0416 and MACS\,J1206), all around $0.13$. Let us notice this study uses three to four potentials across all clusters, and thus our models should be expected to yield larger ratios of core-to-outskirts densities.
Moreover, as this comparison uses $M_{200}$ values from weak-lensing shear-and-magnification analyses \citep[see][]{2014ApJ...795..163U}, we can only cautiously compare it to our X-rays and extrapolated strong lensing measurements.
As the ratio is much higher for MACS\,J0242, this comparison is one more indication that the concentration of mass in the centre of MACS\,J0242 is particularly high relative to its total mass. This is in good agreement with our conclusion of  the cluster being in a cool-core, relaxed dynamical state.

In the case of MACS\,J0949, the cylindrical mass at $R_{10\%} = 205.6_{-25.4}^{+0.00}$ is $M_{\mathrm{2D}} (<R_{10\%}) = (2.07 \pm 0.14) \times 10^{14} \Msol$ using our strong lensing measurements, and with $R_{10\%} = 184.9_{-11.6}^{+12.6}$, $M_{\mathrm{2D}} (<R_{10\%}) = (1.48 \pm 0.05) \times 10^{14} \Msol$ with the X-rays NFW inferred profile. The respective ratios are $0.140 \pm 0.025$ and $0.146 \pm 0.029$. For this cluster again, we notice these ratios to be quite close to one another, supporting the quality of the strong lensing $M_{\Delta}$ extrapolation in spite of the large difference between the X-rays and strong lensing measured masses.
Interestingly, the comparison with the $0.13$ ratio from \citet{2018ApJ...864...98B} hints towards a relative concentration of mass slightly more important in MACS\,J0949.

As we have established through strong lensing models the total matter density distribution in two galaxy clusters, we laid the foundations of our companion paper (Allingham et al. in prep.). In this forthcoming paper, we describe a new method using analytical models of galaxy cluster potentials to predict the ICM distribution, and in the foreseeable future to put constraints on interacting dark matter.

\section*{Acknowledgements}
JA would like to thank Markus Mosbech for comments and discussions.
JA is supported by the International Postgraduate Research Scholarship in Astroparticle Physics/Cosmology at the University of Sydney.
MJ and DJL are supported by the United Kingdom Research and Innovation (UKRI) Future Leaders Fellowship `Using Cosmic Beasts to Uncover the Nature of Dark Matter' (grant number MR/S017216/1). 
DJL is partially supported by ST/T000244/1 and ST/W002612/1.
The authors acknowledge the Sydney Informatics Hub and the use of the University of Sydney high performance computing cluster, Artemis.
This work is based on observations taken by the RELICS Treasury Program (GO 14096) with the NASA/ESA HST, which is operated by the Association of Universities for Research in Astronomy, Inc., under NASA contract NAS5-26555.
GM acknowledges funding from the European Union’s Horizon 2020 research and innovation programme under the Marie Skłodowska-Curie grant agreement No MARACHAS - DLV-896778.
ACE acknowledges support from STFC grant ST/P00541/1.

\section*{Data Availability}

The galaxy and spectroscopic detections catalogues and the lens models are available upon reasonable request to the corresponding author.



\bibliographystyle{mnras}
\bibliography{bibli}




\appendix

\section{Spectroscopic detections of interest}
\label{sec:spectroscopic_detections_of_interest}

We present additional spectroscopic good detections in the background of both clusters MACS J0242 and MACS J0949, respectively in Tables \ref{tab:spectro_mul_m0242_additional} and \ref{tab:spectro_mul_m0949_additional}.
\begin{table}
	\centering
	\caption{Spectroscopic detections of singly imaged objects in MACS\,J0242. Coordinates are in degrees (J2000). The reference for right ascension and declination are taken to be the centre of the cluster.}
	\label{tab:spectro_mul_m0242_additional}
	\begin{tabular}{lccc}
	    \hline
		\hline
		Id. & R.A. & Dec. & $z$\\
		\hline
		10 & 40.6559072 & -21.5412424 & 0.5756\\
        11 & 40.6536287 & -21.5327925 & 0.5928\\
        12 & 40.6546722 & -21.5328188 & 0.5937\\
        13 & 40.6466813 & -21.5480705 & 0.5942\\
        14 & 40.6517158 & -21.5453613 & 0.5943\\
        15 & 40.6566147 & -21.5399484 & 0.5943\\
        16 & 40.6552620 & -21.5388619 & 0.7707\\
        17 & 40.6551537 & -21.5382257 & 0.7713\\
        18 & 40.6407123 & -21.5444971 & 0.8363\\
        19 & 40.6508138 & -21.5463873 & 0.8380\\
        20 & 40.6457745 & -21.5366071 & 3.1120\\
		\hline
		\hline
	\end{tabular}
\end{table}
\begin{table}
	\centering
	\caption{Spectroscopic detections of singly imaged objects images in MACS\,J0949. Coordinates are in degrees (J2000). The reference for R.A. and declination are taken to be the centre of the cluster.}
	\label{tab:spectro_mul_m0949_additional}
	\begin{tabular}{lccc}
	    \hline
		\hline
		Id. & R.A. & Dec. & $z$\\
		\hline
		10 & 147.46989360 & 17.11231290 & 0.5841\\
        11 & 147.46892360 & 17.12212680 & 0.6395\\
        12 & 147.45946988 & 17.11584094 & 0.8472\\
        13 & 147.46832980 & 17.11256280 & 0.8473\\
        14 & 147.46913850 & 17.12435220 & 0.8488\\
		\hline
		\hline
	\end{tabular}
\end{table}
We present in Tables \ref{tab:cl_members_m0242} and \ref{tab:cl_members_m0949} (respectively for clusters MACS\,J0242 and MACS\,J0949) a few cluster members in their final catalogue format: their positions and all geometrical components (semi-major and minor axes $a$ and $b$, rotation angle $\theta$) as well as their magnitudes are coming from the photometric analysis, while the redshifts are detected through spectroscopy.

\begin{table}
	\centering
	\caption{The brightest cluster members in the cluster MACS\,J0242. Coordinates are in degrees (J2000). We remind that the reference coordinates are $(40.649555 ; -21.540485) \deg$. Magnitudes are given on the reference band ACS/F606W. All spectroscopic redshift detections are also provided.}
	\label{tab:cl_members_m0242}
	\resizebox{\hsize}{!}{\begin{tabular}{lccccccc}
	    \hline
		\hline
		Id. & $\Delta_{\alpha}$ & $\Delta_{\delta}$ & $a$ & $b$ & $\theta$ & Mag. & $z$\\ 
		\hline
1 & $0.04387$ & $-0.08964$ & $1.886$ & $1.499$ & $1.83$ & $17.765$ & $0.3130$\\ 
2 & $-31.28771$ & $72.89640$ & $1.027$ & $0.396$ & $2.34$ & $19.898$ & \_ \\ 
3 & $59.25290$ & $79.37028$ & $0.595$ & $0.593$ & $-14.20$ & $20.055$ & \_ \\ 
4 & $82.31906$ & $-5.37408$ & $0.829$ & $0.501$ & $23.90$ & $20.081$ & \_ \\ 
5 & $-47.40417$ & $-5.82480$ & $0.731$ & $0.410$ & $-4.47$ & $20.214$ & \_ \\ 
		\hline
		\hline
	\end{tabular}}
\end{table}

\begin{table}
	\centering
	\caption{Brightest cluster members in the MACS\,J0949. Coordinates are in degrees (J2000). We remind that the reference coordinates are $(\alpha_c, \delta_c) = (147.4659012, 17.1195939)$. Magnitudes are given on the reference band ACS/F814W.}
	\label{tab:cl_members_m0949}
	\resizebox{\hsize}{!}{\begin{tabular}{lccccccc}
	    \hline
		\hline
		Id. & $\Delta_{\alpha}$ & $\Delta_{\delta} $ & $a$ & $b$ & $\theta$ & Mag. & $z$\\
		\hline
1 & $-51.61743$ & $-32.11128$ & $1.344$ & $0.709$ & $45.31$ & $18.761$ & \_ \\ 
2 & $0.05608$ & $-0.15120$ & $1.344$ & $0.740$ & $-57.20$ & $18.789$ & $0.3829$\\ 
3 & $-17.02960$ & $5.76108$ & $0.704$ & $0.657$ & $60.72$ & $18.875$ & $0.3817$\\ 
4 & $51.33490$ & $121.06692$ & $0.742$ & $0.529$ & $50.52$ & $18.970$ & \_ \\ 
5 & $15.93092$ & $-74.92248$ & $0.812$ & $0.526$ & $-24.40$ & $19.054$ & \_ \\ 
		\hline
		\hline
	\end{tabular}}
\end{table}

\section{Additional information on colour-magnitude diagrammes selections}
\label{sec:additonal_CM_CC_data}

We here provide the equation of each main red colour sequence for both galaxy cluster MACS\,J0242 and MACS\,J0949, according to process described in Section \ref{sec:Photo_selection}.
We also provide all the additional colour-magnitude diagrammes we can plot.
Tables \ref{tab:global_eq_CMdiag_m0242} and \ref{tab:global_eq_CMdiag_m0949} provide respectively the equations of the main colour sequences of clusters MACS\,J0242 and MACS\,J0949, and the weighed colour standard deviation of the spectroscopically confirmed cluster galaxy sample $\sigma_C$. The height of the selection box is $2\sigma_{C}$ away from the main red sequence for \textit{HST}/ACS and \textit{HST}/WFC3, and $3\sigma_C$ for \textit{HST}/WFPC2 and DES.

\begin{table}
	\centering
	\caption{Equations of the main colour sequences and standard deviations on colours for all colour-magnitude diagrammes of MACS\,J0242. $m_1$ represents the magnitude in abscissa. Associated graphs are Fig.\,\ref{fig:CM_diag} and \ref{fig:CM_diag_m0242_DES_multiple_v10.17}.}
	\label{tab:global_eq_CMdiag_m0242}
	\begin{tabular}{cccc}
	    \hline
		\hline
		Filter 1 & Filter 2 & $\sigma_C$ & Main colour sequence equation\\
		\hline
		\hline
		& & \textit{HST}/WFPC2 & \\
		\hline
        F814W & F606W & $0.0508$ & $-0.0307 m_1 + 1.6176$\\
        \hline
		\hline
		& & DES & \\
		\hline
		z & r & $0.0466$ & $-0.0382 m_1 + 2.078$ \\
		z & g & $0.1651$ & $-0.0744 m_1 + 4.651$ \\
		r & g & $0.1319$ & $-0.0415 m_1 + 2.779$ \\
		\hline
		\hline
	\end{tabular}
\end{table}

\begin{figure*}
\begin{minipage}{0.1\textwidth}
\end{minipage}
\hfill
\begin{minipage}{0.33\textwidth}
\centerline{\includegraphics[width=1\textwidth]{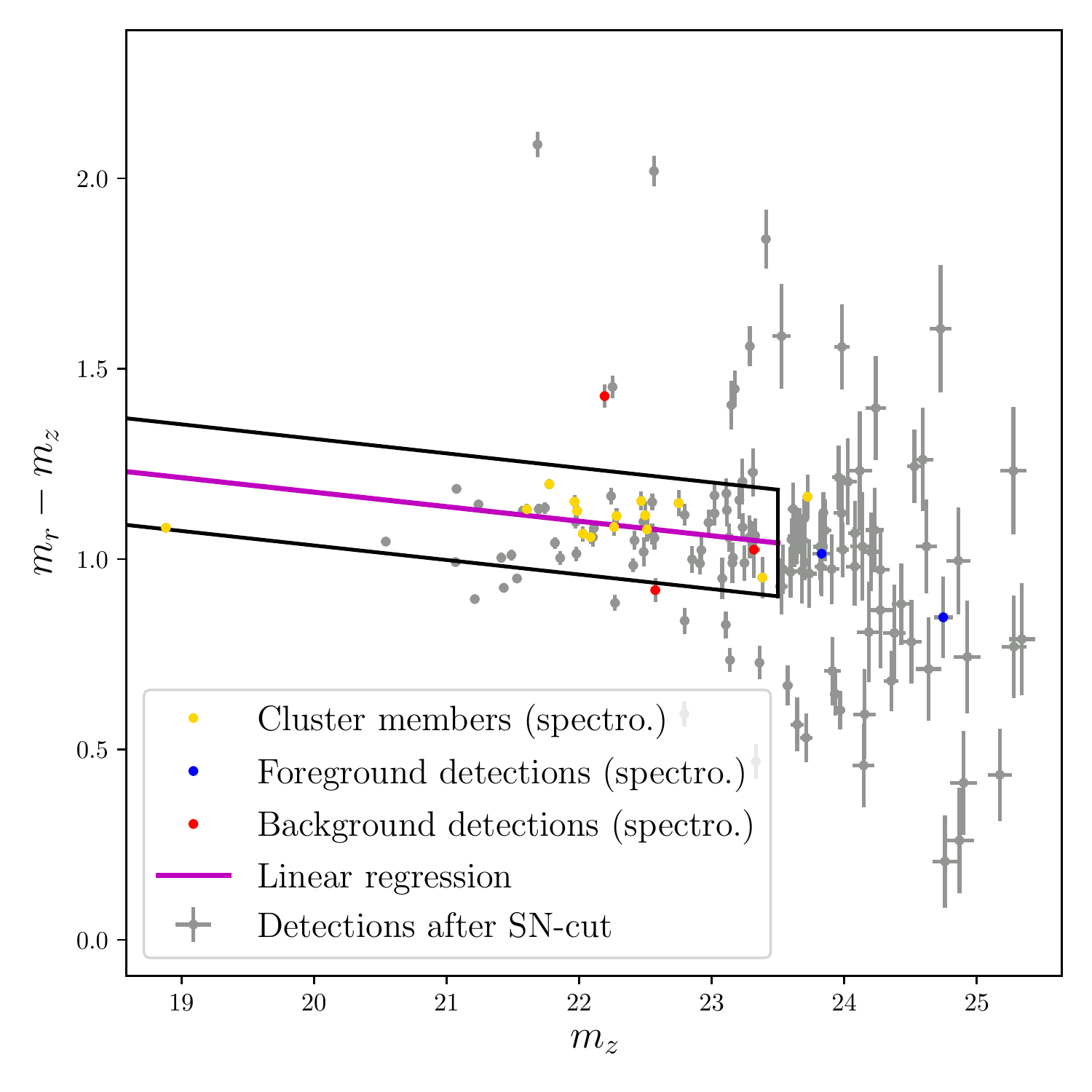}}
\end{minipage}
\hfill
\begin{minipage}{0.33\linewidth}
\centerline{\includegraphics[width=1\textwidth]{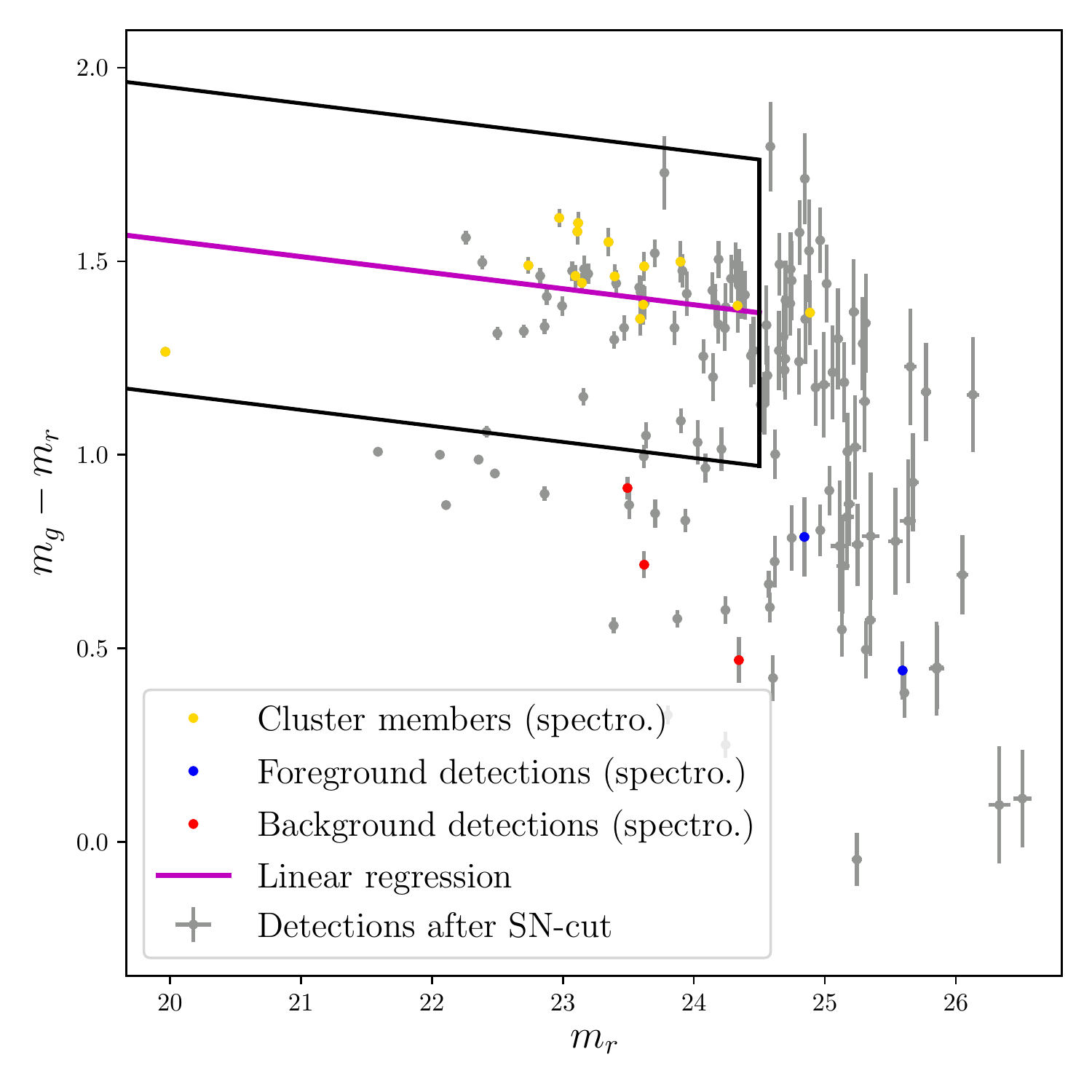}}
\end{minipage}
\hfill
\begin{minipage}{0.1\textwidth}
\end{minipage}
\caption{Colour-magnitude diagramme for MACS\,J0242, instrument DES.
\textit{Top row:} \textit{Left:} The colour is ($m_{\mathrm{r}} - m_{\mathrm{z}}$), and the magnitude $m_{\mathrm{z}}$. \textit{Right:} $m_{\mathrm{z}}$ vs ($m_{\mathrm{g}} - m_{\mathrm{r}}$). 
Grey filled circles (with their error bars) have successfully passed all selections described in Section\,\ref{sec:Extraction}. The magenta line represents the main sequence regression. Blue, gold and red dots represent spectroscopic detections of foreground, cluster and background objects respectively.}
\label{fig:CM_diag_m0242_DES_multiple_v10.17}
\end{figure*}

\begin{table}
	\centering
	\caption{Equations of the main colour sequences and standard deviations on colours for all colour-magnitude diagrammes of MACS\,J0949. $m_1$ represents the magnitude in abscissa. Associated graphs are Fig.\,\ref{fig:CM_diag} and \ref{fig:CM_diag_m0949_WFC3_multiple_v9.2}.}
	\label{tab:global_eq_CMdiag_m0949}
	\begin{tabular}{cccc}
	    \hline
		\hline
		Filter 1 & Filter 2 & $\sigma_C$ & Main colour sequence equation\\
		\hline
		\hline
		& & \textit{HST}/ACS & \\
		\hline
        F814W & F606W & $0.1956$ & $-0.0317 m_1 + 2.0530$\\
		\hline
		\hline
		& & \textit{HST}/WFC3 & \\
		\hline
		F160W & F140W & $0.0230$ & $-0.0121 m_1 + 0.4217$ \\
		F160W & F125W & $0.0365$ & $-0.0253 m_1 + 0.8511$ \\
		F160W & F105W & $0.0684$ & $-0.0483 m_1 + 1.6344$ \\
		F140W & F125W & $0.0220$ & $-0.0158 m_1 + 0.5043$ \\
		F140W & F105W & $0.0523$ & $-0.0361 m_1 + 1.2308$ \\
		F125W & F105W & $0.0371$ & $-0.0209 m_1 + 0.7565$ \\
		\hline
		\hline
	\end{tabular}
\end{table}

\begin{figure*}
\begin{minipage}{0.33\textwidth}
\centerline{\includegraphics[width=1\textwidth]{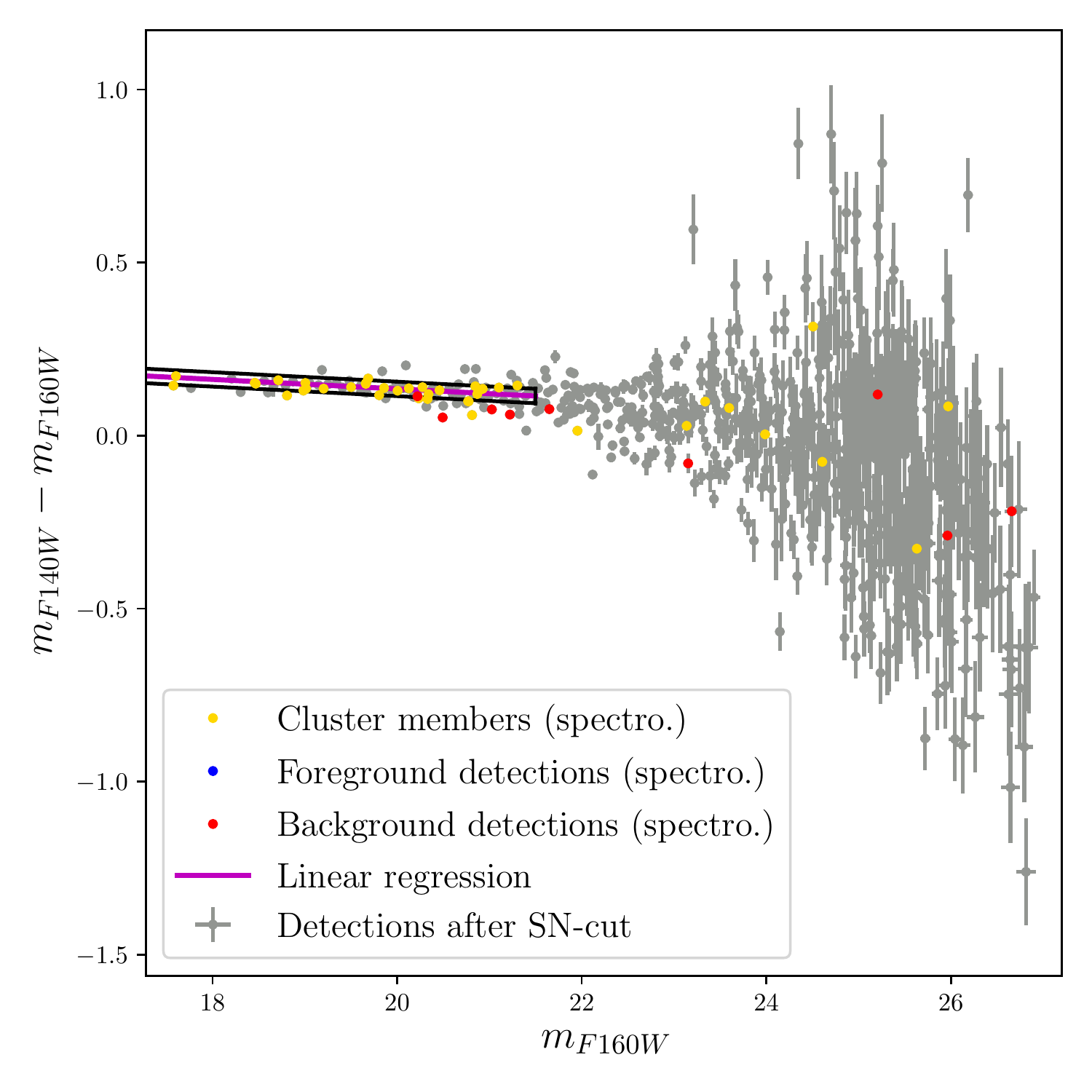}}
\end{minipage}
\hfill
\begin{minipage}{0.33\linewidth}
\centerline{\includegraphics[width=1\textwidth]{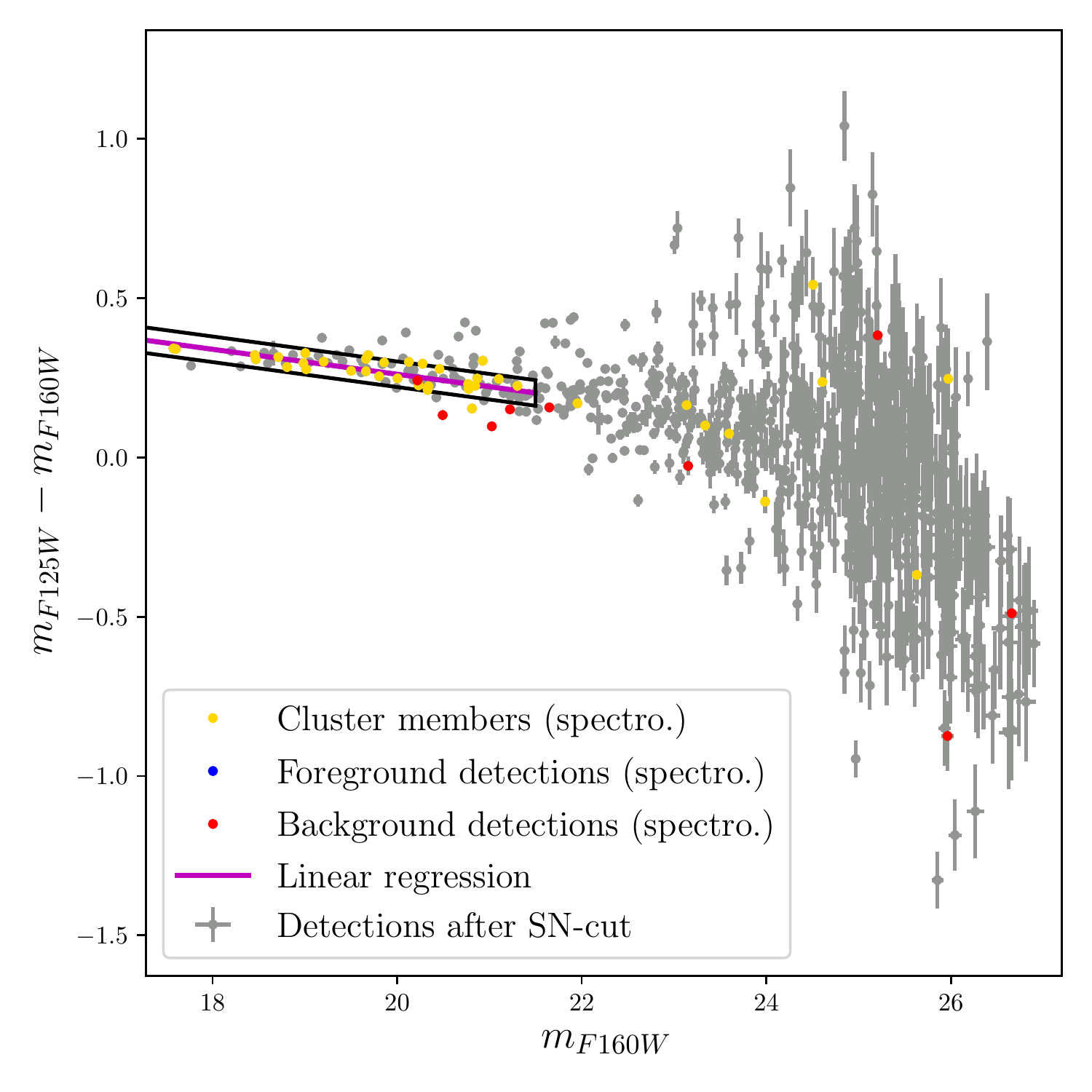}}
\end{minipage}
\hfill
\begin{minipage}{0.33\linewidth}
\centerline{\includegraphics[width=1\textwidth]{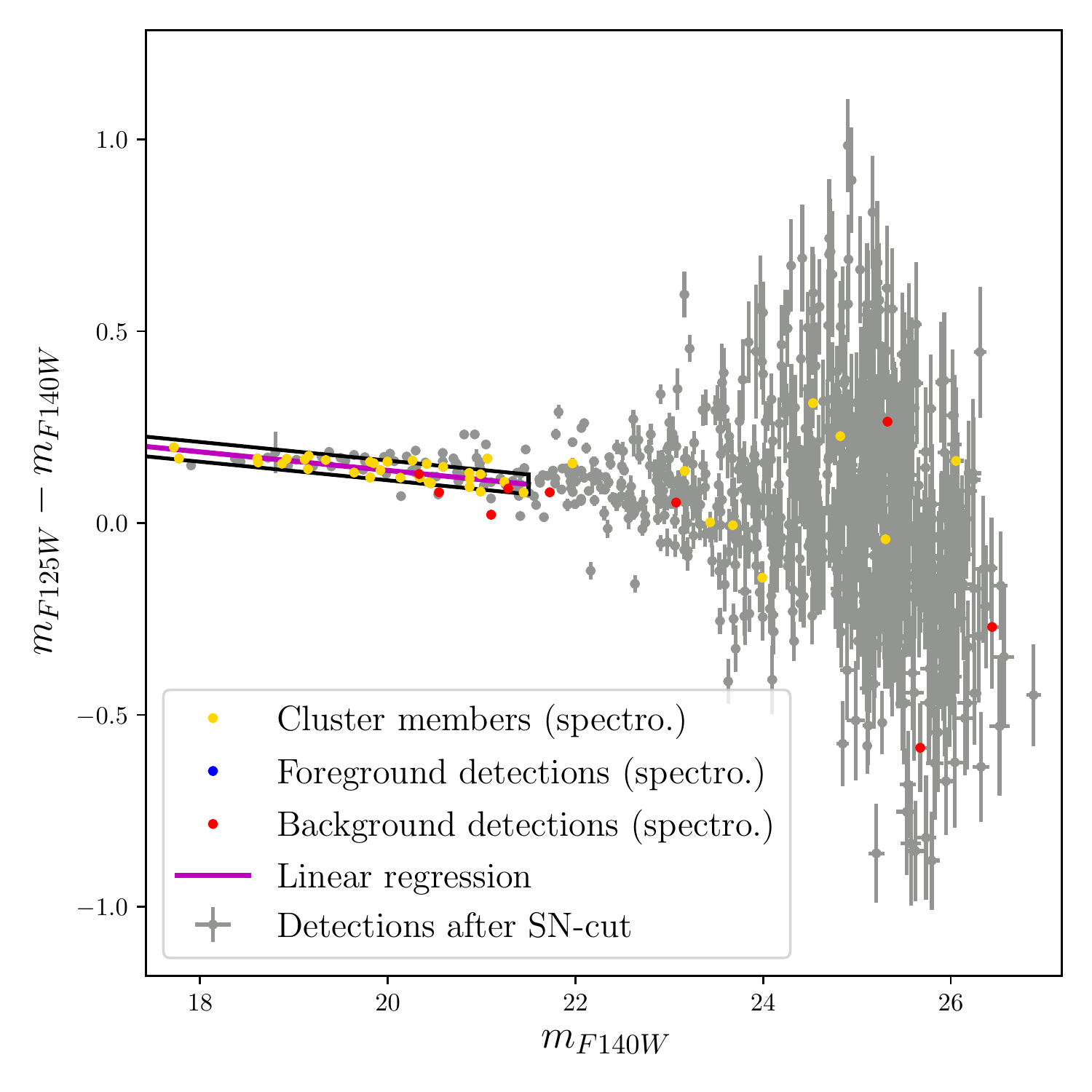}}
\end{minipage}
\begin{minipage}{0.1\textwidth}
\end{minipage}
\hfill
\begin{minipage}{0.33\textwidth}
\centerline{\includegraphics[width=1\textwidth]{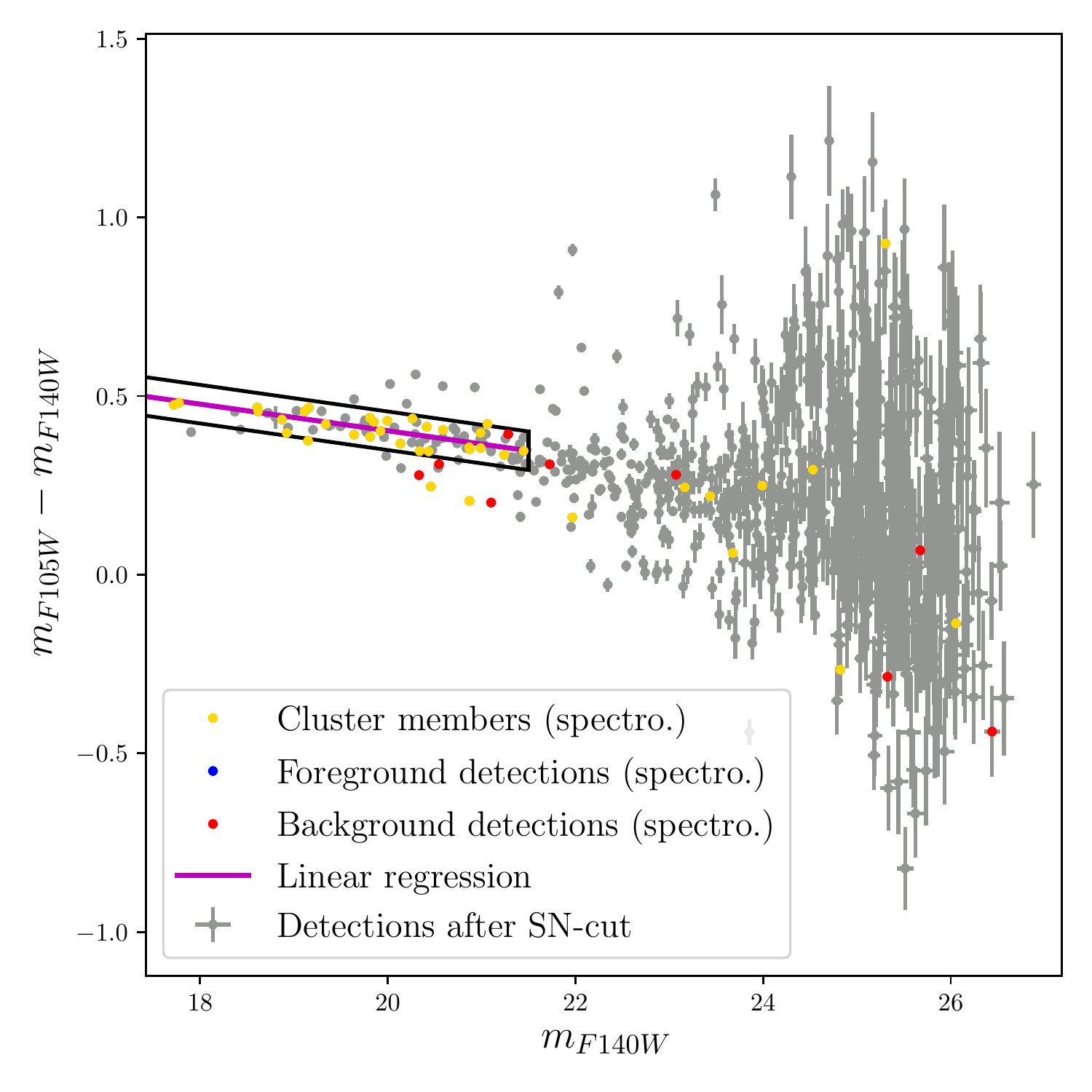}}
\end{minipage}
\hfill
\begin{minipage}{0.33\linewidth}
\centerline{\includegraphics[width=1\textwidth]{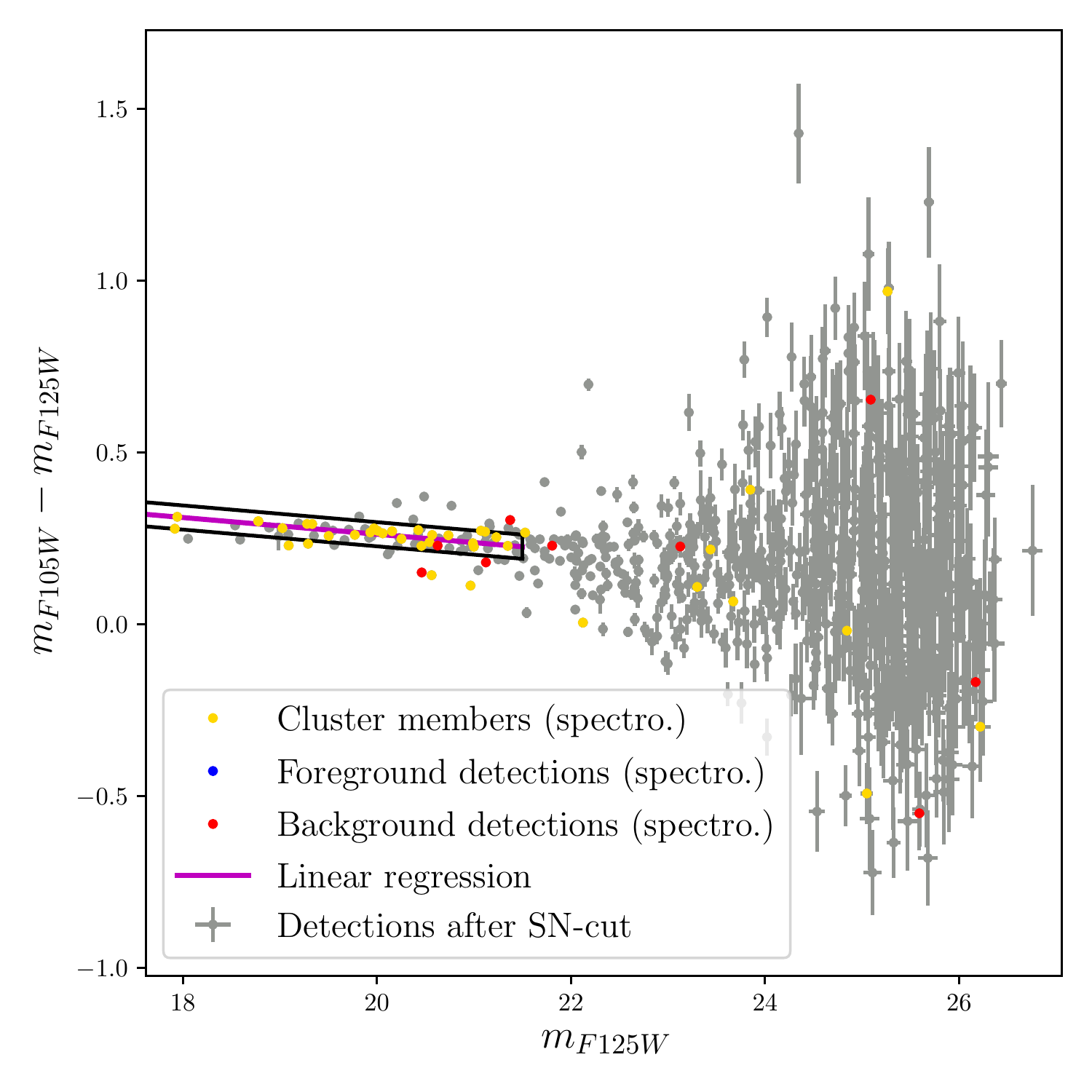}}
\end{minipage}
\hfill
\begin{minipage}{0.1\textwidth}
\end{minipage}
\caption{Colour-magnitude diagrammes for cluster MACS\,J0949, instrument \textit{HST}/WFC3.
\textit{Top row:} \textit{Left:} The colour is ($m_{\mathrm{F140W}} - m_{\mathrm{F160W}}$), and the magnitude $m_{\mathrm{F160W}}$. \textit{Middle:} $m_{\mathrm{F160W}}$ vs ($m_{\mathrm{F125W}} - m_{\mathrm{F160W}}$).  \textit{Right:} $m_{\mathrm{F140W}}$ vs ($m_{\mathrm{F105W}} - m_{\mathrm{F140W}}$).
\textit{Bottom row:} \textit{Left:} $m_{\mathrm{F140W}}$ vs ($m_{\mathrm{F105W}} - m_{\mathrm{F140W}}$). \textit{Right:} $m_{\mathrm{F125W}}$ vs ($m_{\mathrm{F105W}} - m_{\mathrm{F125W}}$).
Grey filled circles (with their error bars) have successfully passed all selections described in Section \ref{sec:Extraction}. The magenta line represents the main sequence regression. Blue, gold and red dots represent spectroscopic detections of foreground, cluster and background objects respectively.}
\label{fig:CM_diag_m0949_WFC3_multiple_v9.2}
\end{figure*}


\bsp	
\label{lastpage}
\end{document}